\documentstyle[12pt]{article}
\newcommand{\be}{\begin{equation}}
\newcommand{\ee}{\end{equation}}
\newcommand{\bea}{\begin{eqnarray}}
\newcommand{\eea}{\end{eqnarray}}
\newcommand{\eps}{\epsilon}
\begin{document}
\title{Phase Dynamics of Bose-Einstein Condensates: Losses versus Revivals}
\author{Alice Sinatra and Yvan Castin \\
Laboratoire Kastler Brossel
\thanks{Unit\'e de recherche de  l'Ecole normale sup\'erieure et de
l'Universit\'e Pierre et Marie Curie, associ\'ee au CNRS.} \\
24 rue Lhomond, 75231 Paris Cedex 5, France}

\date{February 20, 1998}
\maketitle 


\begin{abstract}
In the absence of losses the phase of a Bose-Einstein condensate
undergoes collapses and revivals in time due to elastic atomic interactions.
As experiments necessarily involve inelastic collisions, we develop a model
to describe the phase dynamics of the condensates in presence of collisional
losses. We find that {\it a few inelastic processes} are sufficient to damp
the revivals of the phase.
For this reason the observability of phase
revivals for present experimental conditions
is limited to condensates with a few hundreds of atoms.
\end{abstract}

\section{Introduction}
Since the recent 
experimental observations of Bose-Einstein condensation in dilute
atomic gases \cite{Jila}, \cite{Mit}, \cite{Rice},  
much interest has been raised about the characteristic features of the 
condensate \cite{Parkins}, and about its coherence properties in particular. 
Considerable attention has been devoted to the matter of the relative phase
between two Bose-Einstein Condensates (BECs):
how the phase manifests
itself in an interference experiment (such as the one performed
recently at MIT \cite{interMIT}), how the phase 
can be established by measurement,
and how it evolves in presence of the elastic atomic interactions
(see e.g.\ \cite{Maciek} and references therein).  
In this paper, in view of a possible experimental investigation
of these problems, 
we complete the theoretical work already done on this subject
by studying the dynamics of the relative phase 
in presence of loss processes occurring in the two condensates. 
Such loss processes, unavoidable in a 
real experiment, are due for example to collisions of condensed atoms
with the background gas, or to three-body collisions between condensed atoms 
followed by recombination of two atoms to form a molecule 
\cite{Cornell,Stamper}. 

We consider two mutually non interacting and
spatially non overlapping BECs
in two trapping potentials. We suppose that the experimentalist has at
hand a device, such as the one depicted in fig.\ref{fig:BECs},
allowing both the measurement of the
relative phase between the condensates and the preparation of a state
with a well defined relative phase \cite{CastinDalibard}.
Starting from an initial state with a well defined relative phase, 
we imagine that
the two condensates evolve independently, under the influence of 
the atomic interactions, during a given time interval $t$
at the end of which a measurement of the relative phase is performed. 
By repeating this procedure many times, one accesses the probability 
distribution of the relative phase \cite{CastinDalibard}.

In the lossless case, the relative phase
shows {\it collapses} and {\it revivals} periodically in time 
due to the presence of elastic atomic interactions. In presence of losses,
we find that {\it a few inelastic processes} are sufficient to dramatically
damp the revivals of the phase. In practice, for typical experimental
configurations, the observability of the revivals is limited to
condensates with a small number of atoms, of the order of a 
few hundreds, for which
the revival time is of the order of $0.1$ to $1$ second.
 
\bigskip
In section \ref{sec:Model} we present the theoretical model describing
the evolution of the system in presence of losses. 
An interesting feature of the model is that it can be solved almost exactly
analytically within the Monte Carlo wave function approach
recently put forward by several authors 
\cite{nous,Dum,Gisin,Carmichael}. 
We take advantage of this circumstance in the following sections,
to deduce analytical expressions for the interesting phase-dependent
measurable quantities, and to
a give a simple picture of the phase dynamics in presence of losses: 

In section \ref{sec:SingleReal} we find an approximate
analytical expression for the evolution of a single  
stochastic wave function, and we give a simple physical interpretation
of the result pointing out separately the role of the elastic 
atomic interactions 
and of the losses in the dynamics of the relative phase of the 
condensates. In sections \ref{sec:Observables1} and \ref{sec:Observables2}
we concentrate on the case in which the two condensates are placed in
two identical traps and have initially the same
average number of atoms, and we use the analytical results of section
\ref{sec:SingleReal} to calculate the time dependence of some relative phase 
dependent quantities. In particular
in section \ref{sec:Observables1} we consider an interference
experiment where one counts the atoms detected in the two output channels of 
the beam-splitter of fig.\ref{fig:BECs},
and we analyze the two different physical situations 
in which the condensates' relative phase is 
initially sharply defined or is described by a ``broad'' 
relative phase distribution with a width $\gg \frac{1}{\sqrt{N}}$. 
In section \ref{sec:Observables2} 
we imagine instead an experiment in which the time evolution
of the whole relative phase probability distribution is measured. 
Sections \ref{sec:ExpComp} and \ref{sec:poiss}
are dedicated to the analysis of additional features
that would appear in an experiment;
the effect of asymmetries in the parameters of the two condensates
and in the initial average number of atoms is considered
in section \ref{sec:ExpComp}, and
the effect of fluctuations in the initial total number of atoms
is considered in section \ref{sec:poiss}. 
Some concluding remarks are presented in section \ref{sec:Conclusions}.

\section{The model}
\label{sec:Model}

\subsection{The master equation} 
Let us consider two mutually non-interacting and spatially non-overlapping 
BECs A and B in two harmonic potentials. 
Our starting point to describe the evolution of this system
in presence of $m$-body losses is a master equation for the density matrix $\rho$
describing the atoms in the traps:
\be
\frac{d\rho}{d t} = \frac{1}{i\hbar} [H,\rho]
  + \int d^3 \vec{r} \; \kappa \left[ [\hat{\psi}(\vec{r})]^m \, \rho \, 
     [\hat{\psi}^\dagger(\vec{r})]^m 
  - \frac{1}{2} \{ [\hat{\psi}^\dagger(\vec{r})]^m 
                  [\hat{\psi}(\vec{r})]^m , \rho \} \right] \; 
, \label{eq:uno} 
\ee 
where $\{ X,Y\}$ denotes the anticommutator, and
$[\hat{\psi}^\dagger(\vec{r})]^m$ is the field operator raised to
the power $m$ which suppresses $m$ particles in $\vec{r}$.
In second quantized form the Hamiltonian $H$ reads:
\be
H=\int d^3 \vec{r} \left[ \hat{\psi}^\dagger(\vec{r}) H_0 \hat{\psi}(\vec{r}) +
   \frac{g}{2} \hat{\psi}^\dagger(\vec{r}) \hat{\psi}^\dagger(\vec{r}) 
   \hat{\psi}(\vec{r})  \hat{\psi}(\vec{r}) \right] \; , 
\label{eq:hamilt}
\ee
where $H_0$ is the one-particle Hamiltonian including the trapping 
potential and the kinetic energy, and $g=4 \pi \hbar^2 a/M$ where $M$ is
the mass of the atoms and $a$ is the $s$-wave scattering length. 

The loss terms in Eq.(\ref{eq:uno}) are parameterized by the
number $m$ of particles lost per collisional event and by the collisional 
constant $\kappa$.
Physically the case $m=1$ corresponds to collisions of atoms
in the condensate with atoms of background gas in the cell;
the case $m=2$ corresponds to spin-flip collisions between condensed
atoms in magnetic traps, 
as only specific spin components are trapped; the case $m=3$
corresponds to three-body collisions between condensed atoms, leading
to the formation of an excited molecule and a hot atom supposed to
leave the condensate. The collisional constant $\kappa$ for the processes $m=1$
and $m=3$ has been measured for ${}^{87}$Rb atoms at JILA \cite{Cornell} and 
for ${}^{23}$Na atoms at MIT \cite{Stamper}.
The collisional constant
for the $m=2$ process has not been accurately measured for
these atoms yet, as the two-body losses 
seem to give a smaller contribution to the total decay rate.

We assume that at any time the state of the condensate A (resp.\ B)
can be described in terms of a single occupied mode,
neglecting the excitations out of this mode
due to a non-zero temperature or to the loss processes.
We assume furthermore that these modes are the single particle ground
state wave functions $\phi_a,\phi_b$ given 
self-consistently as functions of the number of particles 
by the Gross-Pitaevskii equation:
\be
\left[ H_0 + gN_\eps |\phi_\eps(\vec{r};{N}_\eps)|^2 \right] 
         \phi_\eps(\vec{r};N_\eps) = \mu_\eps(N_\eps) \phi_\eps(\vec{r};N_\eps) \;,
\label{eq:GPE}
\ee 
where the $\mu_\eps(N_\eps)$'s are the chemical potentials for
the condensates with $N_\eps$ particles, and where the wave functions
$\phi_\eps$ are normalized to unity.
In more mathematical words we approximate the atomic field operator by:
\be
\hat{\psi}(\vec{r})=\sum_{\eps=a,b} c_\eps \phi_\eps(\vec{r};\hat{N}_\eps)
\label{eq:psiat}
\ee
where the operators $c^\dagger_a$ ($c^\dagger_b$) 
and $c_a$ ($c_b$) create and annihilate a 
particle in the condensate A (B) respectively, and where 
$\hat{N}_\eps=c^\dagger_\eps c_\eps$ are the operators giving
the number of particles in each condensate. Note that we keep in
Eq.(\ref{eq:psiat}) the dependence of the mode on the number of 
particles in the condensate. 

By substituting Eq.(\ref{eq:psiat}) into Eq.(\ref{eq:hamilt}) we get
\be
H=E_a(\hat{N}_a)+E_b(\hat{N}_b)
\ee
with
\be
E_\eps(N_\eps)=N_\eps \left[ \int d^3 \vec{r} 
        \phi_\eps^\ast(\vec{r};{N}_\eps) H_0  
        \phi_\eps(\vec{r};{N}_\eps) +
\frac{gN_\eps}{2} |\phi_\eps(\vec{r};{N}_\eps)|^4 \right] 
\label{eq:ener}\;
\ee 
(we have used $N_\eps-1\simeq N_\eps$).

By assuming that in the considered time interval
the atom number distributions in the two condensates remain 
peaked around the initial average values:
\be
\bar{N}_\eps=\mbox{Tr}[\rho(0) c_\eps^\dagger c_\eps] \:,
\ee
we expand the condensates' Hamiltonian  
around $\bar{N}_a$, $\bar{N}_b$ keeping
up to the quadratic terms: 
\be 
H(\hat{N}_a,\hat{N}_b) \simeq H^{q}(\hat{N}_a,\hat{N}_b)   
 \equiv  \sum_{\eps = a,b} E(\bar{N_\eps})+ (\hat{N}_\eps-\bar{N}_\eps) 
           \, \mu_\eps(\bar{N}_\eps) 
+ \frac{1}{2} (\hat{N}_\eps-\bar{N}_\eps)^2  \mu_\eps^\prime(\bar{N_\eps})\; . 
\label{eq:Hquad}
\ee 
In our model we will use this quadratic version of the Hamiltonian, where the 
chemical potentials $\mu_a$ and $\mu_b$  and their derivatives
can be calculated by solving numerically
the Gross-Pitaevskii equation (\ref{eq:GPE}).

We now substitute our ansatz Eq.(\ref{eq:psiat}) in the loss part
of the master equation; since the condensates do not overlap
this amounts to the substitution
\be
[\hat{\psi}(\vec{r})]^m \rightarrow \sum_{\eps=a,b} [\hat{c}_\eps
\phi_\eps(\vec{r};\hat{N}_\eps)]^m
\label{eq:psim}
\ee
in Eq.(\ref{eq:uno}).
In contrast to the Hamiltonian part which required a careful quadratization
in $\hat{N}_\eps -\bar{N}_\eps$
to get the correct phase dynamics,
the dissipative part will be treated
to lowest order by replacing
$\hat{N}_\eps$ by $\bar{N}_\eps$ in Eq.(\ref{eq:psim}).
This allows us finally to obtain a master equation of the form:  
\be 
\frac{d\rho}{d t} = 
  \frac{1}{i\hbar} [H^q(\hat{N}_a,\hat{N}_b),\rho] + 
\sum_{\eps = a,b}\gamma_\eps[c_\eps]^m \rho
 [c_\eps^\dagger]^m  - \frac{\gamma_\eps}{2} \{ [c_\eps^\dagger]^m 
      [c_\eps]^m,  \rho \} \; , 
\label{eq:me}
\ee 
where (for $\eps=a,b$) we have introduced the rates for the 
$m$-body collisions:
\be
\gamma_\eps= \kappa \int d^3\vec{r} 
            |\phi_\eps(\vec{r};\bar{N}_\eps)|^{2m} 
\label{eq:gamma}\; . 
\ee
 
\subsection{Stochastic formulation}
To study the evolution of the system we 
adopt the Monte Carlo wave function point of view 
\cite{nous} which provides us with
a stochastic formulation of the master equation (\ref{eq:me}). 
To this aim we introduce the jump operators:  
\be
S_\eps= \sqrt{{\gamma}_\eps} [c_\eps]^m \hspace{1cm} \eps=a,b 
\ee
and an effective Hamiltonian:
\be
H_{\mbox{\scriptsize eff}}= H^q - \frac{i \hbar}{2} \sum_{\eps=a,b} 
         S_\eps^\dagger S_\eps \:. 
\label{eq:Heff}
\ee  
The Monte Carlo wave function $|\psi(t)\rangle$
undergoes a non hermitian Hamiltonian evolution ruled by
$H_{\mbox{\scriptsize eff}}$ (plus a continuous renormalization)
interrupted by random quantum jumps occurring at a rate
$\langle\psi(t)|\sum_{\eps=a,b} (S_\eps^\dagger S_\eps) |\psi(t)\rangle$,
where $|\psi(t)\rangle$ is normalized to unity. The effect of a quantum
jump is to replace $|\psi\rangle$ by $S_\eps |\psi\rangle$ up to
a normalization factor. Physically this corresponds to the loss
of $m$ particles in the condensate $\eps$ {\sl via} the $m$-body collisional
processes described above.
The two kinds of jumps $\eps=a,b$ occur with relative probabilities:
\be
\frac{P_{a}}{P_{b}}=
\frac{\langle\psi(t)|S_a^\dagger S_a|\psi(t)\rangle}
{\langle\psi(t)|S_b^\dagger S_b|\psi(t)\rangle} \;. 
\label{eq:pjump}
\ee
Starting with a state with a fixed total number of particles $N$,
we can expand at each time the state vector on the Fock basis
\be
|\psi(t)\rangle=\sum_{N_a=0,\tilde{N}} d_{N_a} |N_a,\tilde{N}-N_a\rangle \;, 
\label{eq:fock}
\ee
where $\tilde{N}$ is the total number of atoms at time $t$ in the two 
condensates, and we can carry out the evolution numerically.
The mean value of an observable $\hat{O}$ is obtained by averaging the 
expectation value $\langle\psi(t)|\hat{O}|\psi(t)\rangle$  
over all possible stochastic realizations
for the evolution of $|\psi(t)\rangle$.

Usually the Monte Carlo wave function technique is carried out
purely numerically. It turns out that for the present problem
it is possible to treat analytically the evolution
of a Monte Carlo wave function and, after a minor approximation,
average analytically over all the possible stochastic realizations.
This leads to a simple interpretation of the dynamics
and allows the derivation of analytical formulas for
observables' mean values.
As it will appear in the figures the analytical results are in
good agreement with the numerical results.

\section{Evolution of a single wave function} 
\label{sec:SingleReal}

In this section we derive an approximate formula for the evolution of a 
single stochastic wave function, and we discuss its physical interpretation.  
We first consider the simple case in which the condensates are initially
in a {\it phase state}, introduced in the beginning of the section, and
subsequently the general case in which the initial state is characterized by
a given relative phase distribution. 

For the following it will be useful to introduce the operators 
\be
\hat{N}=\hat{N}_b+\hat{N}_a \hspace{2cm}\hat{n}=\hat{N}_b-\hat{N}_a
\label{eq:nN}
\ee          
corresponding to the sum and difference of the number of atoms in A and in B.
 
\subsection{Phase states}
A very useful class of states of two condensates is represented by the 
{\it phase states} \cite{Leggett}:
\be
|\phi\rangle_N=\frac{1}{\sqrt{2^N N!}} \; (c_a^\dagger e^{i \phi} +
                 c_b^\dagger e^{-i \phi})^N \; |0\rangle
\label{eq:phase}
\ee
having a fixed total number of particles $N$ and leading
to a well defined relative phase $2\phi$ between the condensates
A and B. These states have the remarkable properties:
\bea
c_\eps |\phi
\rangle_N &=& \sqrt{\frac{N}{2}} e^{i\phi \,( \delta_{\eps,a}-
                           \delta_{\eps,b}) } \; |\psi\rangle_{N-1}
\hspace{2cm} \eps=a,b
\label{eq:pro1}  \\
e^{-i \alpha \hat{n}} |\phi\rangle_N &=& |\phi+\alpha\rangle_{N} 
\hspace{4.3cm} \forall \alpha \;,
\label{eq:pro2}
\eea
where the $\delta_{\eps,\eps^\prime}$ for $\eps,\eps^\prime=a,b$ are Kronecker
deltas.
The first property reflects the fact that in a phase state,
all the particles are in the
same state (see Eq.(\ref{eq:phase})), and the second one
shows that $n$ and $\phi$ are  to some extent conjugate variables
like the momentum and position of a particle.
Note that the phase states are not orthogonal:
\be
{}_N \langle\phi^\prime|\phi\rangle_N=[\cos(\phi-\phi^\prime)]^{N} \;,
\label{eq:scalphase}
\ee 
though the function $[\cos(\phi-\phi^\prime)]^N$ in Eq.(\ref{eq:scalphase})
becomes very peaked around zero when $N \rightarrow \infty$ with a width
scaling as $1/\sqrt{N}$.
Any state with a total number $N$ of particles can be expanded on
the overcomplete set of phase states:
\be
|\psi\rangle=
{\cal A} \int_{-\pi/2}^{\pi/2}\; \frac{d\phi}{\pi}\; c(\phi)\; |\phi
\rangle_N \label{eq:expa} \;, 
\ee
where $c(\phi)$ can be obtained from the expansion of the
state vector on the Fock state basis:
\be
c(\phi)= {\cal A}^{-1} \sum_{N_a=0,N}
2^{N/2} \left(\frac{N_a!(N-N_a)!}{N!}\right)^{1/2} 
e^{i(N-2N_a)\phi} \; 
\langle N_a,N-N_a|\psi\rangle \;.
\label{eq:inversion}
\ee
The quantity $|c(\phi)|^2$ can be interpreted as 
the relative phase probability distribution \cite{CastinDalibard}. 
This distribution, flat for a Fock state and very peaked for a phase state,
is normalized in such a way that:
\be
\int_{-\pi/2}^{\pi/2}\; \frac{d\phi}{\pi}\; |c(\phi)|^2 = 1 \;.
\label{eq:normcphi}
\ee
The factor ${\cal A}$ in Eq.(\ref{eq:expa})
ensures that $|\psi\rangle$ is normalized to unity.
For $N \gg 1$ and for a $c(\phi)$ varying slowly at the scale $1/\sqrt{N}$,
we can replace the scalar product ${}_{N}\langle\phi'|\phi\rangle_{N}$
by the delta distribution $\sqrt{2\pi/N}\delta(\phi-\phi')$ to obtain
${\cal A}=\left( \frac{\pi N}{2}\right)^{1/4}$.

\subsection{Approximate expression for $|\psi(t)\rangle$}
Consider the evolution of the state vector $|\psi(t)\rangle$,  
from a time $t_0=0$ to a time $t$, for {\it a particular stochastic realization}. 
We imagine that $k$ quantum jumps, each corresponding to the loss of
$m$ particles, occur at times $t_1,...,t_k$ separated
by time intervals $\tau_j=t_j-t_{j-1}$ with $j=1,...,k$; the $k^{th}$ 
jump takes place in the condensate $\eps_k$ with $\eps_k=a,b$. We have:    
\be
|\psi(t)\rangle={\cal N} e^{-\frac{i}{\hbar} H_{\mbox{\scriptsize eff}}
                                             (t-t_k)} S_{\eps_k}
      e^{-\frac{i}{\hbar} H_{\mbox{\scriptsize eff}}\tau_k} S_{\eps_{k-1}}
    e^{-\frac{i}{\hbar} H_{\mbox{\scriptsize eff}}\tau_{k-1}} ...\,S_{\eps_1} 
    e^{-\frac{i}{\hbar} H_{\mbox{\scriptsize eff}}\tau_{1}} |\psi(0)\rangle 
\label{eq:start}
\ee
where ${\cal N}$ is a normalization factor.
By using the identity:
\be
[c_\eps]^m f(\hat{N_a},\hat{N_b}) =
    f(\hat{N_a}+m \delta_{\eps,a},\hat{N_b}+m \delta_{\eps,a})\, [c_\eps]^m 
    \hspace{1cm} \eps=a,b  \:,
\label{eq:shifta}
\ee
we shift all the jump operators in Eq.(\ref{eq:start}) to the right by 
letting them ``pass through'' the exponentials and we obtain: 
\bea
|\psi(t)
\rangle&=&{\cal N} \exp[-i H_{\mbox{\scriptsize eff}}
                                       (\{\hat{N}_\eps\})(t-t_k)/\hbar] 
  \exp[-iH_{\mbox{\scriptsize eff}}
  (\{\hat{N}_\eps+m \delta_{\eps,\eps_k}\})\tau_k/\hbar]  \nonumber \\
  &\hspace{0.2cm}&    \exp[-iH_{\mbox{\scriptsize eff}}
                  (\{\hat{N}_\eps+m (\delta_{\eps,\eps_k}+
                  \delta_{\eps,\eps_{k-1}}\})\tau_{k-1}/\hbar] \,
              ... \prod_{j=1,k} S_{\eps_j} |\psi(0)\rangle
\label{eq:dotdot}\; .
\eea
We introduce now the major approximation in our calculations by replacing
$[c_\eps^\dagger]^m [c_\eps]^m$ by $\bar{N_\eps}^m$
in the expression for the effective Hamiltonian Eq.(\ref{eq:Heff}),
supposing that the fraction of lost particles is small. The resulting
effective Hamiltonian then takes the form:
\be
H_{\mbox{\scriptsize eff}}= H^q - \frac{i \hbar}{2} \lambda 
\label{eq:Heffana} \: ,
\ee  
quadratic in $\hat{N}_a$ and $\hat{N}_b$, where $\lambda$
is a constant representing
the mean total number of collisional events per unit of time:
\be
\lambda = \lambda_a + \lambda_b \hspace{1cm} \mbox{with} \hspace{1cm}
\lambda_a=\gamma_a \bar{N_a}^m \;, \hspace{0.5cm}
\lambda_b=\gamma_b \bar{N_b}^m\;.
\label{eq:lambda}
\ee
In this approximation 
the statistics of the quantum jumps is simply Poissonian with a parameter
$\lambda$ and $\delta_{b,\eps_j}=1-\delta_{a,\eps_j}$ takes the values $1$ and 
$0$ with probabilities $\lambda_b/\lambda$ and $\lambda_a/\lambda$
respectively, according to  Eq.(\ref{eq:pjump}).

We then expand the effective Hamiltonians in each exponential 
in  Eq.(\ref{eq:dotdot}) around $\hat{N}_a$, $\hat{N}_b$ in powers of
$m \delta_{\eps,\eps_k}$, $m (\delta_{\eps,\eps_k}+\delta_{\eps,\eps_{k-1}})$,
etc. Due to the quadratic dependence of  Eq.(\ref{eq:Heffana}) on 
$\hat{N}_a$ and $\hat{N}_b$
we limit the expansion at the first order, the subsequent terms being
constants or zero. By using Eq.(\ref{eq:Heffana}) we then obtain the following result 
for the state vector at time $t$:
\be
|\psi(t)
\rangle={\cal N} e^{-\lambda t/2}
U_0(t) U_1(t) \prod_{j=1,k} S_{\eps_j}  |\psi(0)\rangle\;.
\label{eq:res}
\ee
In Eq.(\ref{eq:res}) we have introduced the unitary operators
\bea
U_0(t) &=& \exp[-iH^q(\{\hat{N}_\eps\}) t/\hbar] \\
U_1(t) &=&
\exp\left[-i \left(
\frac{\partial H^q}{\partial {N}_a}(\{\hat{N}_\eps\})\Delta_a
+\frac{\partial H^q}{\partial {N}_b}(\{\hat{N}_\eps\})\Delta_b
\right)/\hbar\right]
\eea
where for $\eps=a,b$:
\be
\Delta_\eps = m \sum_{j=1,k} \sum_{l=j,k}   \delta_{\eps,\eps_l} \tau_j =
                       m \sum_{l=1,k} \delta_{\eps,\eps_l} t_l 
\label{eq:Deltae} 
\ee
are random quantities that depend on the particular realization. 

We sketch out briefly the physical interpretation of the  
result  Eq.(\ref{eq:res}), 
considering the action of the successive factors in  Eq.(\ref{eq:res}) 
on a {\it phase state} defined in Eq.(\ref{eq:phase}). 
\begin{itemize}
\item{
The factor $U_0(t)$ in  Eq.(\ref{eq:res})
accounts for the evolution in absence of losses.
Expressed in terms of the operators $\hat{N}$ and $\hat{n}$ 
of Eq.(\ref{eq:nN}) it involves:  
\be
H^q(\{N_\eps\})= 
                 f_0(\hat{N})+\hat{n}v(\hat{N})+ 
                 \hat{n}^2 (\mu^\prime_b+\mu^\prime_a)/8 \; .
\label{eq:hpre}
\ee
We have used  Eq.(\ref{eq:Hquad}) and we have defined
\be
v(\hat{N})={1\over 2\hbar}\{\mu_b-\mu_a
+ {\mu'_b-\mu'_a \over 2} (\hat{N}-\bar{N}) 
- {\mu'_b+\mu'_a \over 2} (\bar{N}_b-\bar{N}_a) \}
\label{eq:v} \; , 
\ee
where $\bar{N}=\bar{N}_a+\bar{N}_b$ and where $\mu_\eps$ stands for 
$\mu_\eps(\bar{N}_\eps)$.
From the properties of the phase state we find that the terms in 
$\hat{n}$ and $\hat{n}^2$ in  Eq.(\ref{eq:hpre}), when exponentiated in $U_0$,
(i) shift the relative phase at the $N$-dependent constant speed $v(\hat{N})$
and (ii) spread the relative phase (in a way analogous to the spreading 
of a wave packet of a massive particle under free evolution), respectively. 
The term $f_0(\hat{N})$ in  Eq.(\ref{eq:hpre}) is a function 
of the total number of atoms $N$ only and plays no role, since it amounts
in $U_0(t)$ to adding
a global phase factor to the wave function. The phase-spreading will 
eventually lead to a {\it collapse} of the relative phase \cite{Parkins}. 
On the other hand due to the discreteness of     
the spectrum of the operator $\hat{n}$ (the spectrum of $\hat{n}$ consists 
of even integers for an even $N$, and of odd integers for an odd 
$N$), 
there are special times at which the exponential operator  Eq.(\ref{eq:hpre}) 
reduces to a mere translation of the relative phase, 
yielding the well known result that {\it revivals} should follow the 
collapses of the relative phase. 
More precisely if one uses the expansion  Eq.(\ref{eq:fock}) 
for the phase state defined in Eq.(\ref{eq:phase}), one realizes
that a relative phase distribution initially peaked around $\phi_0$ 
displays revivals at the times:
\be
t_R=q \pi /\chi\;, \hspace{2cm} q \; \mbox{integer}
\label{eq:trev}
\ee
where we have introduced:
\be
\chi=\frac{\mu_a^\prime+\mu_b^\prime}{2\hbar} \label{eq:chi} \;.
\ee
At these times, for $N$ even: 
\be
e^{-i \chi \hat{n}^2 t_R/4} |\phi\rangle_N = |\phi+ q \pi /2\rangle_N 
\label{eq:even}
\ee
and for $N$ odd:
\be
e^{-i \chi \hat{n}^2 t_R/4} |\phi\rangle_N = e^{-iq \pi/4}  |\phi\rangle_N 
\label{eq:odd}
\ee
The initial relative phase distribution is then reconstructed
around $(\phi_0+v(N)t_R+q\pi/2)$ for $N$ even
and around $(\phi_0+v(N)t_R)$ for $N$ odd.}  
\item{
The factor $U_1(t)$ in  Eq.(\ref{eq:res}) accounts for the 
presence of losses. Expressed in terms of the operators $\hat{n}$
and $\hat{N}$, it involves:
\be
\frac{\partial H^q}{\partial {N}_a}(\{N_\eps\}) \, \Delta_a/\hbar
+\frac{\partial H^q}{\partial {N}_b}(\{N_\eps\}) \, \Delta_b/\hbar
=f_1(\hat{N})+ \hat{n} D
\label{eq:rtrasl} 
\ee
where global phase factors are included in $f_1(\hat{N})$.
The translation operator $\hat{n}$ appears in  Eq.(\ref{eq:rtrasl}) 
multiplied by a random quantity $D$ defined as: 
\be
D = m \sum_{l=1,k} t_l \left[ \chi \delta_{b,\eps_l} - 
              \frac{\mu^\prime_a}{2\hbar} \right] \;.
\label{eq:Delta}
\ee

Equations  (\ref{eq:pro2}) and  (\ref{eq:rtrasl}) show that
the relative phase in a single stochastic realization is shifted by the 
random amount $D$ due to the loss processes. 
This effect will turn out to have a dramatic influence on
the coherence properties of the condensates.} 
\item{
Finally in Eq.(\ref{eq:res}) the action of the jump operators on a phase 
state is simply: 
\be
\prod_{j=1,k} S_{\eps_j} |\phi\rangle_n= \left[ \frac{N}{2} 
    \frac{N-1}{2} \ldots \frac{N-mk+1}{2} \right]^{1/2} 
    e^{-i \phi \alpha} |\phi\rangle_{N-mk}   
\label{eq:effjump}
\ee
where we have introduced the quantity
\be
\alpha= m \sum_{j=1,k} \left[ 2 \delta_{b,\eps_j}-1 \right] \; .
\ee 
Apart from numerical factors that will be absorbed in
the normalization and the phase factor involving $\alpha$,
Eq.(\ref{eq:effjump}) amounts to reducing by a random amount the total 
number of particles.} 
\end{itemize}
In the general case, an initial state with $N$ particles
can be expanded on the phase states set (see  Eq.(\ref{eq:expa})).
By using Eqs. (\ref{eq:hpre}), (\ref{eq:rtrasl}),
(\ref{eq:effjump}), and getting rid of the global phase factors 
we then obtain the wave function: 
\be
|\psi(t)\rangle=  {\cal B}(t) \int_{-\pi/2}^{\pi/2} \frac{d\phi}{\pi} c(\phi,0) 
e^{-i\chi \hat{n}^2 t/4} \; e^{-i \phi \alpha} 
|\phi+D+v(N-mk)t\rangle_{N-mk}  
\label{eq:respsit} \;, 
\ee
where ${\cal B}(t)$ is a normalization factor.

\section{Mean beating intensity of the condensates}
\label{sec:Observables1}

To monitor the evolution of the relative phase between the condensates, a
possible choice is to determine the relative phase dependent
quantity $\langle c_a^\dagger c_b\rangle$
after some time during which the two condensates, initially prepared in a
state with a defined relative phase, evolve independently. 
As the relative phase between the condensates is affected by the
elastic atomic interactions, the average $\langle c_a^\dagger c_b\rangle$ undergoes
collapses and revivals in time.

In the situation described in fig.\ref{fig:BECs} the measure of 
$\langle c_a^\dagger c_b\rangle$ would correspond to
the following measurement scheme: $\;$ Prepare a state in which A and B
have a well defined relative phase \cite{CastinDalibard}; let the condensates 
evolve during a time interval $t$; then let $p \ll N$ atoms escape from the 
condensates and beat them on the beam-splitter.
The counts registered in the two output channels of the beam-splitter  
will be fluctuating 
variables whose averages over many realizations of the whole 
procedure are \cite{CastinDalibard}: 
\be 
I_\pm =
   \langle \frac{p}{\hat{N}}
           \frac{(c_a^\dagger\pm c_b^\dagger)(c_a\pm c_b)}{2}
           \, \rangle \simeq \frac{p}{\bar{N}} 
           \frac{1}{2}\left(\langle c_a^\dagger c_a\rangle
+\langle c_b^\dagger c_b\rangle     \pm
        2 \mbox{Re} \langle c_a^\dagger c_b \rangle \right)\;,
\ee
The difference between $I_+$ and $I_-$ gives then the real part of
$\langle c_a^\dagger c_b\rangle$. 

\bigskip
We shall now use the approximated formulas (\ref{eq:res}) and
(\ref{eq:respsit}) to calculate the time dependence of 
$\langle c_a^\dagger c_b\rangle$.
The main result of this section is that the revivals in this quantity
are damped in time with a simple exponential law
$e^{-\lambda t}$ where the constant $\lambda$, defined in
Eq.(\ref{eq:lambda}), is the mean number of loss processes per unit
of time.

In the present and in the following section we restrict for simplicity to the
perfectly {\it symmetric} case where the two trapping potentials are identical
and the two condensates have initially the same mean number of particles:
\begin{eqnarray}
\bar{N}_a &=& \bar{N}_b  \label{eq:symm1} \;,\\
\gamma_a &=& \gamma_b \label{eq:symm2} \;,\\
\mu_a &=& \mu_b \;.\label{eq:symm3}
\end{eqnarray}
Moreover we consider an initial state having a {\it fixed} total number
of particles equal to $N$; and as a reminder of this choice (when it is the case)
we will attach a superscript $\langle ... \rangle^{\mbox{\scriptsize fix}}$ 
to the averages.
The {\it non symmetric} case for the condensates will be considered in
section \ref{sec:ExpComp}; while the effect of fluctuations in the
initial total number of atoms (requiring a further averaging over $N$) 
will be analyzed in section \ref{sec:poiss}.

We calculate $\langle c_a^\dagger c_b\rangle^{\mbox{\scriptsize fix}}$ 
in two different physical
situations. The first one refers to a sharply defined initial relative phase  
($\Delta \phi \simeq \frac{1}{\sqrt{N}}$) 
for which we choose a {\it phase state} as the initial 
state; the second one, probably more realistic from the experimental point 
of view, makes use of an initial phase
distribution much broader than $\frac{1}{\sqrt{N}}$. In each case we 
first calculate the expectation value of the operator 
$\hat{O}=c_a^\dagger c_b$ 
for a single stochastic realization using the results of section
\ref{sec:SingleReal}, and then take the average over the stochastic
realizations. In the whole paper we will
denote with $\langle \psi(t)| \hat{O} |\psi(t)\rangle$ the single
realization expectation value and with $\langle \hat{O} \rangle$ the
quantum mechanical average. 

\subsection{Case of an initial phase state}
Let us assume $|\psi(0)\rangle=|\phi\rangle_N$;  
by using equations (\ref{eq:res}) and
(\ref{eq:hpre}), (\ref{eq:rtrasl}), (\ref{eq:effjump}), 
for a single realization, we find: 
\be
\langle \psi(t)| c_a^\dagger c_b|\psi(t)\rangle=
  {}_{N-mk} \langle\phi+D|
  e^{ i \frac{\chi}{4} \hat{n}^2 t} c_a^\dagger c_b 
  e^{- i \frac{\chi}{4} \hat{n}^2 t} |\phi+D\rangle_{N-mk} 
  \label{eq:manipul} 
\ee 
where
$\chi$ and $D$ are defined in 
Eq.(\ref{eq:chi}) and  Eq.(\ref{eq:Delta}) respectively. 
Note that the contribution involving the drift velocity of Eq.(\ref{eq:v})
vanishes as we are considering here the {\it symmetric} case.
The quadratic dependence on $\hat{n}$ 
in  Eq.(\ref{eq:manipul}) can be
eliminated by shifting $c_a^\dagger c_b$ through the exponential
$e^{- i \frac{\chi}{4} \hat{n}^2 t}$ using Eq.(\ref{eq:shifta}):
\be
e^{ i \frac{\chi}{4} \hat{n}^2 t} c_a^\dagger c_b e^{- i \frac{\chi}{4} \hat{n}^2 t}
=
e^{-i\chi (\hat{n}+1) t} c_a^\dagger c_b
\label{eq:shift}
\ee
so that
\be
\langle \psi(t)| c_a^\dagger c_b |\psi(t) \rangle= 
   {}_{N-mk}\langle\phi+D|
      e^{-i\chi (\hat{n}+1) t} c_a^\dagger c_b
      |\phi+D\rangle_{N-mk} \; ; 
\ee
by using the properties  (\ref{eq:pro1}), (\ref{eq:pro2}) and
(\ref{eq:scalphase}) we then have: 
\be
\langle \psi(t)|c_a^\dagger c_b|\psi(t)\rangle=
     {\frac{N-mk}{2}} \; 
     e^{-2 i \phi} e^{-2i D} [\cos(\chi t)]^{N-mk-1} \;. 
\label{eq:moynda}
\ee 
The next step is to take the average of the result  Eq.(\ref{eq:moynda})
over the stochastic realizations which amounts to averaging  
over the random variables $k$, $\tau_j$ and $\delta_{b,\eps_j}$ (the last two variables 
appearing in the random quantity $D$). We show the calculation of
the average in detail 
in the appendix A. The result for $\langle c_a^\dagger c_b 
\rangle^{\mbox{\scriptsize fix}}$ reads: 
\be
\langle c_a^\dagger c_b \rangle^{\mbox{\scriptsize fix}}=
   e^{-2i\phi} e^{-\lambda t}
  \sum_{k=0,N/m-1} \frac{N-mk}{2}
         \frac{1}{k!}
         \left[ \lambda t\; u(t) \right]^k 
          \left[ \cos(\chi t) \right]^{N-mk-1}
\label{eq:aveapp} \;, 
\ee
where the function $u(t)$ is given by:
\be
u(t)={\sin(m \chi t)\over m\chi t} \;. 
\label{eq:usymm}
\ee
By identifying the factor $N-mk$ with $N$ under the assumption of a small fraction of lost particles, and by extending the sum over $k$ up to $\infty$,
we are able to express the result in a compact way:
\footnote{It should be noted however that the compact formula 
(\ref{eq:resacrb}) diverges for $\chi t=\pi/2+q\pi$,
where the explicit sum  Eq.(\ref{eq:aveapp}) should be used instead. 
At such points $\langle c_a^\dagger c_b \rangle^{\mbox{\scriptsize fix}}=0$ anyway.}
\be
\langle c_a^\dagger c_b \rangle^{\mbox{\scriptsize fix}}=
   e^{-2i\phi} e^{-\lambda t[1-u(t)/\cos^m(\chi t)]} {\frac{N}{2}}
          [\cos(\chi t)]^{N-1} \;. 
\label{eq:resacrb}
\ee
The factor $[\cos(\chi t)]^{N-1}$ in  Eq.(\ref{eq:resacrb}),
already obtained in \cite{WongCollett} in the absence of losses,
is responsible for the 
collapses of the average value $\langle c_a^\dagger c_b\rangle^{\mbox{\scriptsize fix}}$ 
and for revivals at times $t_R=q \pi/\chi$ with $q$ integer. 
The collapses and revivals of $\langle c_a^\dagger c_b\rangle^{\mbox{\scriptsize fix}}$
are shown in fig.\ref{fig:acrb} both (a) in absence and (b) in presence
of three-body losses.
We see immediately that the losses have a dramatic effect reducing
exponentially in time the average with  
the rate $\lambda$ given by  Eq.(\ref{eq:lambda}). In fact
at a revival times $t=t_R$, $u(t)$ vanishes so that
the average value of $\langle c_a^\dagger c_b\rangle^{\mbox{\scriptsize fix}}$
is simply attenuated with respect to the lossless case: 
\be
\langle c_a^\dagger c_b\rangle_{t=t_R}^{\mbox{\scriptsize fix}}=
(-1)^{q(N-1)} \langle c_a^\dagger c_b\rangle_{t=0}^{\mbox{\scriptsize fix}} 
\; e^{-\lambda t_R}\;, 
\label{eq:acrbtrev}
\ee 
by an exponential factor which is exactly 
the probability that no particles are lost up to time $t$.
The dramatic effect of losses on the revivals, 
already when $\lambda t_R\simeq 1$
(that is {\it one} loss process has occurred on average at the revival time),
can be understood by the fact that in each single realization experiencing
a quantum jump at a time $t\sim t_R$ the relative phase is shifted by an
amount $D \sim \pi$. This point will be further exemplified in 
section \ref{sec:Observables2}.

\subsection{Case of an initial relative phase distribution broader
than that of a phase state} \label{sub:broader}
Since it may be difficult to prepare experimentally the condensates
in a phase state we now consider the more realistic
case in which the initial relative phase
distribution $|c(\phi,0)|^2$ for the condensates is broad as compared to
$1/\sqrt{N}$. To be specific we assume that the initial
relative phase distribution is a Gaussian centered at $\phi=0$:
\be  
c(\phi,0)=
{\cal G}_0 \exp\left(-\phi^2/(4 \Delta \phi^2) \right) 
\hspace{2cm} \frac{1}{\sqrt{N}} \ll\Delta \phi \ll 1  \:, 
\label{eq:gauss}
\ee
where $\phi$ ranges between $-\pi/2$ and $\pi/2$.
This choice corresponds to a Gaussian distribution for the number
of particles in the condensates:
\be
\langle N_a, N-N_a|\psi(0)\rangle =
 {\cal G} e^{-(N-2N_a)^2/4 \Delta n^2}
\ee
with $\Delta n \, \Delta \phi=1/2$.

For a single realization, we use  Eq.(\ref{eq:respsit}) and
we proceed along the lines of the previous calculation to get:
\bea
\langle \psi(t) |c_a^\dagger c_b|\psi(t)\rangle =
{\left[ \frac{\pi \tilde{N}}{2}  \right]^{1/2}} \int_{-\pi/2}^{\pi/2} 
&&\frac{d\phi}{\pi} \frac{d\phi^\prime}{\pi} c(\phi,0) c^\ast(\phi^\prime,0)
{\frac{\tilde{N}}{2}} e^{-i\alpha(\phi-\phi^\prime)} \nonumber\\
  && e^{-i(\phi+\phi^\prime+2D)} 
{}_{\tilde{N}-1}\langle\phi^\prime-\chi t|\phi\rangle_{\tilde{N}-1}\; 
\eea
where $\tilde{N}=N-mk$ with $k$ equal to the number of
quantum jumps experienced by the Monte Carlo wave function up to time $t$.
Now by using the fact that 
the scalar product between the phase states for $N\gg 1$
is a very peaked function of $\phi-\phi'$ 
with respect to 
the other functions in the integral, we perform the substitution:
\be
{}_{\tilde{N}-1} \langle \phi^\prime-\chi t|\phi\rangle_{\tilde{N}-1}
\rightarrow
\cos^{\tilde{N}-1}(q_0\pi)\sqrt{\frac{2\pi}{\tilde{N}}}\delta(\phi^\prime+q_0\pi-\chi t-\phi)
\ee
where the integer $q_0$ is chosen 
such that $-\pi/2<(\chi t+\phi-q_0\pi)\leq \pi/2$.
As the factor $c(\phi,0)$ defined in 
Eq.(\ref{eq:gauss})
is peaked around $\phi=0$, we neglect the dependence of $q_0$ on $\phi$
so that the integer $q_0$ is finally chosen such that 
$-\pi/2<(\chi t-q_0\pi)\leq \pi/2$.
In this way we obtain
\bea
\langle \psi(t)| c_a^\dagger c_b |\psi(t)\rangle &=&
(-1)^{q_0(N-1)} \frac{\tilde{N}}{2}
e^{i(\chi \alpha t-2D)} \nonumber\\
&&\int_{-\pi/2}^{\pi/2} \frac{d\phi}{\pi}  
c(\phi,0) c^\ast(\phi+\chi t-q_0\pi,0) \, e^{-i(2\phi+\chi t-q_0\pi)} \;. 
\label{eq:bof}
\eea
The next step is to average the factor 
$e^{i(\chi \alpha t-2D)}$ over the 
stochastic realizations. The procedure closely follows the one in the
appendix A.
By identifying $\tilde{N}$ with $N$, 
as in the previous case, and by extending the
boundaries of integration in Eq.(\ref{eq:bof}) to $\pm\infty$ we can express
the result in the compact form\footnote{
To obtain  Eq.(\ref{eq:Yvan}) we use the condition $\Delta \phi<<1$ 
to set:
\be
\langle c_a c_b^\dagger \rangle_{t=0}^{\mbox{\scriptsize fix}} = \frac{N}{2}
 \left( \int_{-\pi/2}^{\pi/2} \frac{d\phi}{\pi}
            c^2(\phi,0) \, e^{-2i\phi} \right) \simeq \frac{N}{2} \;.
\ee}: 
\be
\langle c_a^\dagger c_b\rangle^{\mbox{\scriptsize fix}}=
{ \frac{N}{2} } e^{-\lambda t[1-u(t)]}
 \sum_{q=0}^{+\infty} e^{-[(\chi t-q \pi)/2]^2 /2\Delta \phi^2} 
 (-1)^{q(N-1)}   
\label{eq:Yvan}
\ee
where $u(t)$ is defined in Eq.(\ref{eq:usymm}).
The factor involving the sum over $q$ 
in  Eq.(\ref{eq:Yvan}) plays the role of the factor 
$[\cos(\chi t)]^{N-1}$ in Eq.(\ref{eq:resacrb}) which was obtained 
for an initial phase state. 
At each time $t_R=q\pi/\chi$ there is a 
revival of the quantity $\langle c_a^\dagger c_b\rangle^{\mbox{\scriptsize fix}}$ and 
Eq.(\ref{eq:Yvan}) reduces to the  very simple expression:
\be
\langle c_a^\dagger c_b\rangle_{t=t_R}^{\mbox{\scriptsize fix}}=
(-1)^{q(N-1)} \langle c_a^\dagger c_b\rangle_{t=0}^{\mbox{\scriptsize fix}}
 \; e^{-\lambda t_R}
\label{eq:Yvantrev}
\;. 
\ee
This formula does not depend on the initial width $\Delta\phi$
and coincides with the one Eq.(\ref{eq:acrbtrev}) obtained for
a phase state. There is therefore no possible way of reducing the
damping of the revivals by adjusting the initial width
of the phase distribution. Only the temporal width of
the revivals is larger for a distribution broader 
than that for a phase state, as it clearly appears from a comparison
between fig.\ref{fig:acrb2} and the previous fig.\ref{fig:acrb}b.

\bigskip
{\it Remark}: Formula (\ref{eq:Yvan}) can also be used to study the collapse
of the phase around $t=0$. For short times ($t\ll t_R$)
we expand $u(t)$ to second
order in $t$ obtaining:
\be
\langle c_a^\dagger c_b\rangle^{\mbox{\scriptsize fix}} \simeq {N\over 2} 
\exp\left\{-{(\chi t)^2\over 8\Delta\phi^2} \left[1+{4\over 3} m^2\Delta\phi^2
\lambda t\right]\right\}
\ee
In the absence of losses we recover the collapse time $t_c=2\Delta\phi/\chi$
\cite{Maciek} as the half temporal width at the relative height $e^{-1/2}$
of the mean signal
$\langle c_a^\dagger c_b\rangle^{\mbox{\scriptsize fix}}$. 
Losses start {\it accelerating} the collapse significantly
when $m^2\Delta\phi^2\lambda t_c> 1$. In this regime of course 
the subsequent revivals cannot be observed.

\section{Evolution of the relative phase distribution}
\label{sec:Observables2}

We turn now our attention to the phase distribution $|c(\phi)|^2$ 
which could be reconstructed in an experiment for example via a 
series of multichannel measurements. We show  an example of the procedure 
in fig.\ref{fig:multich} \cite{JavaninenYoo}, \cite{CastinDalibard}. 

In the frame of our model, the evolution of $c(\phi)$ can
be obtained numerically from the evolution of the state vector 
$|\psi(t)\rangle$ expanded on the Fock state basis by using 
Eq.(\ref{eq:inversion}); however, as we show in the following, 
the approximated analytical treatment allows us also in this case to find some 
simple results at the revival times. 

Let the initial state of the condensate, with a total number $N$ of atoms,
be characterized by a given
relative phase distribution $c(\phi,0)$; the state vector at
time $t$ is then given by our approximated formula Eq.(\ref{eq:respsit}). 
One can easily check that the integrand in Eq.(\ref{eq:respsit}) is periodic
of period $\pi$ 
so that we can shift the interval of integration to obtain: 
\footnote{When $\phi \rightarrow \phi+\pi$, $c(\phi,0)$
is multiplied by $(-1)^N$, $\exp(-i\alpha \phi)$ is multiplied by $(-1)^{mk}$,
and the phase state $|\phi+D+v t\rangle_{\tilde{N}}$ 
is multiplied by $(-1)^{N-mk}$.} 
\be
|\psi(t)\rangle={\cal B}(t)
  e^{-i \chi \hat{n}^2 t/4}
  \int_{-\pi/2}^{\pi/2}\; \frac{d\phi}{\pi}\; 
  \tilde{c}(\phi-D-v(\tilde{N})t,0) |\phi\rangle_{\tilde{N}}   
\label{eq:psitrasl}
\ee 
where $\tilde{c}(\phi)=e^{-i\alpha \phi} c(\phi)$ and $\tilde{N}=N-mk$. 
This result has a very suggestive interpretation: $\;$ The loss processes in a  
single realization {\it shift} the relative phase distribution 
by a random amount $D$,
and the overall evolution can be separated in a random shift plus the
Hamiltonian evolution. To make clearer this interpretation, we have plotted  
in fig.\ref{fig:singreal} the phase distribution at the second revival time
(given by Eq.(\ref{eq:trev}) with $q=2$) for different realizations. 
For $\lambda t_R\simeq 1$, as in the figure,
there is an important fraction of realizations in which the relative phase
is shifted considerably. This is the reason why the relative phase distribution
at the revival time will be smeared out by the losses when we take the 
average over the stochastic realization, which we do now.

As in section \ref{sec:Observables1} we consider the symmetric case defined
by the Eqs.(\ref{eq:symm1}), (\ref{eq:symm2}), (\ref{eq:symm3}). Furthermore we 
restrict ourselves to the revival times $t=t_R
=q\pi/\chi$, $q$ integer (see Eq.(\ref{eq:trev})). 
In this case 
the Hamiltonian evolution operator in Eq.(\ref{eq:psitrasl}) takes a simple
numerical form (see Eq.(\ref{eq:even}) and Eq.(\ref{eq:odd})) 
and by comparing Eq.(\ref{eq:psitrasl})
to Eq.(\ref{eq:expa}) we can simply read out the phase distribution
amplitude $c(\phi,t)$:
\bea  
c(\phi,t_R)&=& \tilde{c}(\phi_{\tilde{N}}-D,0) \;,
\label{eq:cjoli}
\eea
where:
\bea
\phi_{\tilde{N}}&=&\phi-q\pi/2   \hspace{1.cm} \mbox{for}\ \tilde{N}\ \mbox{even} \\
\phi_{\tilde{N}}&=&\phi       \hspace{2.5cm} \mbox{for}\  \tilde{N}\ \mbox{odd}.
\eea
From Eq.(\ref{eq:cjoli}) we see again that a single loss event
(which can lead to $D\sim\pi$) has a dramatic effect on the
phase distribution.

As shown in the appendix B the phase distribution at the revival times
averaged over the stochastic realizations takes the very simple form:
\be 
\langle |c(\phi,t_R)|^2 \rangle^{\mbox{\scriptsize fix}}
= (1 - e^{\lambda t_R}) + e^{- \lambda t_R} |c(\phi_{N},0)|^2 \;. 
\label{eq:suggestive}
\ee 
At the revival time the relative phase distribution is ``damped'' by 
the factor $e^{-\lambda t_R}$ while a flat background component appears.
This effect is clearly shown in fig.\ref{fig:cphirev}, where we have plot
the averaged relative phase distribution at $t=0$ and at the second
revival time.

\section{Effect of an asymmetry between the two condensates}
\label{sec:ExpComp}

In the previous sections we have investigated the relative phase dynamics
in the {\it symmetric} case for the two condensates. In this section
we extend the analysis to account for a small imbalance in the initial
average number of particles 
\be
|\bar{N}_b-\bar{N}_a| \ll \bar{N} \;,
\ee
where $\bar{N}$ is the average of the total initial number of particles,
and for arbitrary values of the parameters $\mu_a$, $\mu_b$,
$\gamma_a$, $\gamma_b$. 
We restrict the calculation to
the contrast of the interference fringes between the two
condensates averaged over many experimental realizations,
assuming an initial phase distribution broader than the phase state.

Our initial Monte Carlo wave function has
a fixed total number of particles equal to $N$,
and a Gaussian distribution for number of particles in each condensate.
The calculation of $\langle c_a^\dagger c_b\rangle^{\mbox{\scriptsize fix}}$
is now slightly more involved than in the symmetric
case, as the phase distribution amplitude $c(\phi,0)$ acquires
a phase factor varying rapidly with $\phi$ at the scale
$1/\sqrt{N}$. All the calculations are therefore put in the appendix C,
and we give here the result only at the revival time $t=t_R$:
\be
\langle c_a^\dagger c_b \rangle_{t=t_R}^{\mbox{\scriptsize fix}} = (-1)^{q(N-1)}
\frac{N}{2} e^{-2 i v(N) t_R} e^{-\lambda t_R[1-U(t_R)]} \;,
\label{eq:resasymtr}
\ee
where $v(N)$ is defined by Eq.(\ref{eq:v}) and $U(t)$ is a function of time
(see Eq.(\ref{eq:newU}) in appendix C).
In fig.\ref{fig:asymm} we show an example of the
time evolution of $\langle c_a^\dagger c_b\rangle^{\mbox{\scriptsize fix}}$ 
in the case of a $10\%$ asymmetry
in the initial number of particles $\bar{N}_a$ and $\bar{N}_b$.
As far as the {\it damping} of the revivals is concerned, no significant
difference appears with respect to the symmetric case. The damping of the
revivals is in this case ruled by the exponent:
\be
-\lambda t_R [1-\mbox{Re}U(t_R)]
\ee
where:
\be
\mbox{Re}U(t_R)= \frac{1}{\lambda}
\left( \, \lambda_b  \; \mbox{sinc}(m \mu_b^\prime t_R/\hbar) +
          \lambda_a  \; \mbox{sinc}(m \mu_a^\prime t_R/\hbar)
\, \right) \;,
\label{eq:RenewU}
\ee
where $\mbox{sinc}(x)=\sin(x)/x$.
Obviously $|\mbox{Re}U(t_R)|\leq 1$, meaning that an asymmetry
between the condensates cannot {\it amplify} the revivals
with respect to the lossless case.
From Eq.(\ref{eq:RenewU}) we notice,
just as a curiosity, that a complete suppression of the effect of the
losses ($\mbox{Re}U(t_R)=1$) would occur only in the case in which there are
no losses in the condensate A ($\lambda_a=0$) and no elastic
interactions in the condensate B ($\mu_b^\prime=0$) (or vice versa).

A trivial effect of the asymmetry, evident in fig.\ref{fig:asymm},
is the appearance of {\it oscillations}
of the mean value $\langle c_a^{\dagger}c_b \rangle^{\mbox{\scriptsize fix}}$ 
due to the non zero drift velocity of the relative phase
of the condensates. We will see in the next section that this effect,
harmless at first sight, can have dramatic consequences
when we consider the effect of the dispersion in the 
initial total number of particles $N$. 

\section{Effect of fluctuations in the total number of particles}
\label{sec:poiss}

Through all the previous sections in this paper
we have chosen an initial state, represented by our initial Monte Carlo wave function,
with a fixed total number of particles in the condensates.
The averages that we calculated 
$\langle ... \rangle^{\mbox{\scriptsize fix}}$ then correspond
to the real quantum mechanical averages
supposing that the initial total number of atoms is
fixed to a value $N$ for any realization of the experiment.
In practice it is probably difficult to produce a Fock state
for the condensates and the total number of atoms should be governed by some
probability distribution $P(N)$.  Since we have analytical formulas
for the quantities of interest (such as the average $\langle c_a^\dagger
c_b \rangle^{\mbox{\scriptsize fix}}$), it is very simple to add a further 
averaging over $N$ for a given $P(N)$. Suppose for example
that the distribution for the initial total number of atoms
is a Poissonian distribution of parameter $\bar{N}$. 
By averaging the result Eq.(\ref{eq:resasymtr}), valid 
at the revival times $t_R$ for slightly asymmetric condensates,
we get:
\be
|\langle c_a^\dagger c_b \rangle_{t=t_R}^{\mbox{\scriptsize Poiss}}| 
= \frac{\bar{N}}{2}
 e^{-\lambda t_R[1-\mbox{\scriptsize Re}U(t_R)]} \,
 e^{-\bar{N} \left\{ \sin^2(\mu_a^\prime t_R/2\hbar)
                     + \sin^2(\mu_b^\prime t_R/2\hbar) \right\} }\;.
\label{eq:fright}
\ee
The result Eq.(\ref{eq:fright}) shows that a slight asymmetry between
the condensates kills the  revivals of $\langle c_a^\dagger c_b \rangle$.
This is due to the fact that the drift
velocity of the relative phase $v(N)$ in Eq.(\ref{eq:resasymtr})
depends on the initial total number of particles,
giving to $\langle c_a^\dagger c_b \rangle_{t=t_R}^{\mbox{\scriptsize fix}}$
a phase factor of the form:
\be
\exp[-2 i v(N)t_R] \propto
\exp[i \, (N-\bar{N}) \, \frac{\mu'_b-\mu'_a}{2\hbar} \, t_R] =
\exp[i \, (N-\bar{N}) \, \frac{\mu'_b-\mu'_a}{\mu'_b+\mu'_a} \, q \pi] \;.
\ee
To be able to observe the revivals it is then necessary to be
as close as possible to the symmetric conditions in order to satisfy:
\be
\frac{\mu'_b-\mu'_a}{\mu'_b+\mu'_a} \Delta N \ll 1 \;,
\label{eq:condiz}
\ee
where $\Delta N$ is the width of the distribution $P(N)$.

If the {\it symmetry} between the condensates is perfectly realized,
the atom number fluctuations have the simple effect of doubling the
revival time.
We show an example in fig.\ref{fig:Poisson} where we averaged the
result for $\langle c_a^\dagger c_b \rangle^{\mbox{\scriptsize fix}}$ for an initial phase state (fig.\ref{fig:acrb}) 
using a Poissonian distribution for $P(N)$. 
The main effect is the
disappearance of the ``odd'' revivals; this is due to
the fact that the amplitude of these odd revivals for $N$ particles
is proportional to $[\cos(q \pi)]^{(N-1)}=(-1)^{(N-1)}$ which alternates
its sign depending on the parity of $N$.

\bigskip
In fact it is possible to show that a Poissonian ensemble of phase states
is equivalent to a {\it coherent state} for the two condensates, as long
as one calculates the mean values of operators commuting with the
total number of particles in the condensates. For the perfectly symmetric
case in fig.\ref{fig:Poisson} we then recover the result
obtained in \cite{WongCollett} (in the absence of losses) i.e.\ the
doubling of the revival period for a coherent state of the condensates
as compared to the phase state.

Within the coherent states pictures we can also reinterpret the
result Eq.(\ref{eq:fright}) for the asymmetric case in the following way:
in order to observe a revival of the relative phase between two
condensates it is necessary that {\it both} condensates display a phase
revival at the same time i.e.\ $\mu_a^\prime/2\hbar t_R=q \pi$ and
$\mu_b^\prime/2\hbar t_R=q^\prime \pi$, with $q,q'$ integers.

\section{Concluding remarks}
\label{sec:Conclusions}
We have studied the dynamics of the relative phase between two 
Bose-Einstein condensates in presence of $m$-body loss processes in order
to question the observability of the {\it collapses} and {\it revivals} 
of the phase predicted by purely Hamiltonian models.

We have shown that the losses damp exponentially 
in time the phase dependent quantity $\langle c_a^\dagger c_b \rangle$
(see Eq.(\ref{eq:acrbtrev}) for an initial phase state
and Eq.(\ref{eq:Yvantrev}) for an initially broader phase distribution).
The decay rate $\lambda$ of $\langle c_a^\dagger c_b \rangle$ coincides 
(up to the factor $m$) with the mean total number of particles  
lost per unit of time, and it is therefore approximately $N$ times larger 
than the inverse lifetime of a particle in the condensates, where $N$
is the total number of particles initially in the condensates.

The dramatic effect of the losses on the relative phase has been
suggestively interpreted within the Monte Carlo wave function approach. 
In a single realization
each single loss event occurring at a time of the order of the revival
time shifts the relative phase by a random amount of the order of $\pi$.
A few loss processes are then sufficient  
to smear out the relative phase completely at the revival time
when the average over the stochastic realizations is taken.   
For this reason the experimental observation of the revivals is limited
to condensates with a small number of atoms
where the condition $\lambda t_R < 1$ (where $t_R$ is the revival time 
Eq.(\ref{eq:trev})) can be satisfied for all the relevant loss
processes in the system. 

In order to give an idea of the possible scenarios and of the order of
magnitudes in different experimental conditions, we have shown 
in fig.\ref{fig:feasib} the loss rates due to one-body and three-body 
collisions
and the inverse revival time as functions of the total
number of atoms, for two different values of the trap frequencies. 
For higher trap frequencies (fig.\ref{fig:feasib}a)
the revivals occur on a shorter time scale
and one is confronted mainly to three-body losses, while for less confining
traps (fig.\ref{fig:feasib}b)
collisions with the residual gas should be taken into account due
to longer revival times.
Fig.\ref{fig:feasib} shows 
that phase revivals in presence of losses are
in principle observable in condensates with some hundreds of atoms.

By studying the general case of two asymmetric condensates, and the
effects of fluctuations in the initial total number of atoms in
the condensates, we have finally pointed out a practical difficulty
which should be overcome in order to observe phase revivals.
The difficulty comes from
the fact that in the case of two non perfectly symmetric condensates
their relative phase drifts with a velocity {\it depending} on the
initial total number of atoms.
For this reason random fluctuations in the initial number of atoms
turn out to destroy the relative phase revivals when the asymmetry
is too large.
A possible way to overcome this problem is of course to use two almost
symmetric condensates. Another possibility, which we have not examined
in detail, would be to use a condensate A which has a collapse time
longer than the duration of the experiment 
($\bar{N}(\mu^\prime_a t_R/\hbar)^2 \ll 1$)
as a {\it phase reference} to measure the evolving phase of the other
condensate B.

\section*{Acknowledgments}
We wish to thank Jean Dalibard for useful discussions, 
and Christopher Herzog for comments on the manuscript. 
Y.C. wishes to thank Keith Burnett for pointing out 
the problem of the influence of losses on the revivals at several 
conferences. A.S. acknowledges financial support from the Atomic Coherence
TMR network ERB FMRX-CT96-0002 of the European Community.

\section*{Appendix A: average of the phase factor $\lowercase{e}^{-2iD}$}
In this appendix we derive the average over the stochastic realizations
of the quantity $e^{-2iD}S(k)$ where $D$ is defined in Eq.(\ref{eq:Delta})
and where $S(k)$ is an arbitrary function of the number of jumps $k$.
We perform the average over the variables $\delta_{b,\eps_j}$ first, using their
probability distribution given after Eq.(\ref{eq:lambda}); we have:
\be
\langle e^{-2 i D }\rangle_{\delta_{b,\eps_j}}=
 \prod_{j=1,k} \frac{1}{\lambda} 
( \lambda_b e^{-\frac{i}{\hbar} m \mu^\prime_b t_j}+
  \lambda_a e^{\frac{i}{\hbar} m \mu^\prime_a t_j} )
\equiv \prod_{j=1,k} f(t_j)
\; .
\label{eq:prod}
\ee
In order to perform the average over the variables $k$ and $\tau_j$, we need the
probability distribution $P_t(k,t_1,t_2,...t_k)$ of having in the time
interval $(0,t)$ exactly $k$
jumps separated by time intervals $\tau_j=t_j-t_{j-1}$. Since we assume that
the loss processes occur randomly with a constant rate $\lambda$,
corresponding to a waiting-time distribution
of the form $w(\tau)=\lambda e^{-\lambda \tau}$, the
probability distribution $P_t(k,t_1,t_2,...t_k)$ is simply 
\cite{Carmichael}:
\be
P_t(k,t_1,t_2,....t_k)=\lambda^k e^{-\lambda t}.
\label{eq:Pkappa}
\ee
Using this result we are led to the calculation of a multiple integral
of the form:
\be
I=\int_{0<t_1<t_2...<t_k<t} f(t_1)f(t_2)...f(t_k) \; dt_1 dt_2 ... dt_k 
\label{eq:integral}
\ee
where $f(t)$ is the argument of the product in Eq.(\ref{eq:prod}).
Since $I$ is equal to $I_\sigma$ calculated for any permutation
$t_{\sigma(1)},...t_{\sigma(k)}$ of the integration variables, we can write 
it as:
\be
I= \frac{1}{k!} \left[ \sum_{\sigma}
  \int_{0<t_{\sigma(1)}<...<t_{\sigma(k)}<t} f(t_1)f(t_2)...f(t_k)
       \; dt_1 dt_2 ... dt_k \right] =
   \frac{1}{k!} \left[ \int_0^t f(t) \; dt \right]^k \;.
\label{eq:cadeau}
\ee
We then obtain
\be
\langle S(k)e^{-2iD}\rangle_{k,\tau_j,\delta_{b,\epsilon_j}}
= \sum_{k\geq 0} S(k) \frac{\lambda^k}{k!} 
  \left[ \int_0^t f(t) \; dt \right]^k e^{-\lambda t}.
\ee
In this last equation we may have to introduce by hand a
cut-off $N/m-1$ over the index $k$ if  $S(k)$ has divergences
for $k\ge N/m$ (i.e. when no particles are left in the condensates).

\section*{Appendix B: phase distribution at revival times}
We are interested in calculating the phase distribution
at the revival time averaged over the realizations
that is $\langle|c(\phi,t_R)|^2\rangle_{k,\tau_j,\delta_{b,\eps_j}}$.
We restrict to the symmetric case between the condensates and
we start from Eq.(\ref{eq:cjoli}).
By using Eq.(\ref{eq:inversion}) for $t=0$ we have: 
\be
\langle|c(\phi,t_R)|^2\rangle_{k,\tau_j,\delta_{b,\eps_j}}= |{\cal A}(0)|^{-2}
\sum_{N_a=0,N} \sum_{N'_a=0,N} \mbox{fac}(N_a) \mbox{fac}^\ast(N'_a)   
  \langle e^{2i(N'_a-N_a) (\phi_{\tilde{N}}-D)}    
  \rangle_{k,\tau_j,\delta_{b,\eps_j}}
\label{eq:prelim}
\ee
where we have introduced the notation
\be
\mbox{fac}(N_a) =
2^{N/2} \left(\frac{N_a!(N-N_a)!}{N!}\right)^{1/2} 
\langle N_a,N-N_a|\psi(0)\rangle \;.
\ee
The calculation of the average over the stochastic realizations closely
resembles the previous one
Eq.(\ref{eq:prod}) that we have explained in the appendix A; we have:
\be
\langle e^{2i(N'_a-N_a)(\phi_{\tilde{N}}-D)} \rangle_{k,\tau_j,
\delta_{b,\eps_j}}
= \sum_{k\geq 0} e^{-\lambda t_R} \frac{(\lambda t_R)^k}{k!} 
\left[ \frac{\sin[(N'_a-N_a)m\chi t_R]}
   {(N'_a-N_a)m\chi t_R} \right]^k
e^{2i(N'_a-N_a)\phi_{\tilde{N}}}\;. 
\label{eq:fatto}
\ee
We note that the terms in the sum in Eq.(\ref{eq:fatto}) for $k\neq 0$
are equal to zero unless $(N'_a-N_a)=0$ in which case the average
in Eq.(\ref{eq:fatto}) is equal to one. We can then rewrite the result (\ref{eq:prelim})
as:
\bea
\langle|c(\phi,t_R)|^2\rangle^{\mbox{\scriptsize fix}} &=& 
 |{\cal A}(0)|^{-2} \left[ \sum_{N_a=0,N} \sum_{N'_a=0,N} 
\delta_{N'_a,N_a}  |\mbox{fac}(N_a)|^2  \right.
 \nonumber \\  
& &
\left. \phantom{ \sum_{N_a=0,N}} +
(1-\delta_{N'_a,N_a}) \left( \mbox{fac}(N_a) [\mbox{fac}(N'_a)]^\ast  
e^{2i(N'_a-N_a) \phi_N} e^{-\lambda t_R} \right) \right] 
 \;. 
\eea 
Now by using the property:
\be
\sum_{N_a=0,N} |\mbox{fac}(N_a)|^2 |{\cal A}(0)|^{-2} =1  
\ee 
coming from the normalization condition Eq.(\ref{eq:normcphi}) and from 
Eq.(\ref{eq:inversion}), we find the suggestive result Eq.(\ref{eq:suggestive}). 

\section*{Appendix C: asymmetric condensates}
In this appendix we show the explicit calculation of the mean contrast
of the interference fringes $\langle c_a^\dagger c_b\rangle^{\mbox{\scriptsize fix}}$
for asymmetric condensates.
We consider an initial Monte Carlo wave function for which 
the total number of particles $N$ is fixed and the 
number of particles in condensate A has a Gaussian probability
distribution:
\be
\langle N_a, N-N_a|\psi(0)\rangle = {\cal G}
  e^{-(N_a-x_a N)^2/\Delta n^2}
\label{eq:gaussn}
\ee
where $\cal G$ is a normalization factor and
$\Delta n$ is the standard deviation for the difference $n$ in the
number of particles in the two condensates.
The quantities $x_a=\bar{N}_a/(\bar{N}_a+\bar{N}_b)$ 
and $x_b=\bar{N}_b/(\bar{N}_a+\bar{N}_b)$ are the average fractions of
particles initially in the condensate A and B respectively,
which are supposed to be fixed 
from one realization to the other even in presence of
fluctuations of the initial total number of atoms.

We suppose in what follows that
\be
1 \ll \Delta n \ll \sqrt{N}\; ,
\label{eq:condn}
\ee
and
\be
|x_a N- x_b N| \ll N \;.
\label{eq:condsigma}
\ee

We first derive the phase distribution amplitude corresponding to
the initial state Eq.(\ref{eq:gaussn})
by using Eq.(\ref{eq:inversion}).  
We evaluate the factorials in Eq.(\ref{eq:inversion}) using the Stirling's formula,
and we use a local approximation valid for $|N_a - x_a N| \ll \sqrt{N}$:
\be
{N_a!(N-N_a)!\over N!} \simeq  {(x_a N)!(x_b N)!\over N!}
e^{(N_a-x_a N)\ln(x_a/x_b)} \;.
\ee
By approximating the discrete sum in Eq.(\ref{eq:inversion})
with an integral over $N_a$ ranging from
$-\infty$ to $+\infty$ we obtain:
\be
c(\phi,0)={\cal N} e^{-\phi^2 \Delta n^2 } e^{i\kappa \phi}
\ee
where:
\be
\kappa=(x_b-x_a) N - {1 \over 2}  \Delta n^2 \ln(x_a/x_b)
\ee
and where ${\cal N}$ is a normalization factor obtained from Eq.(\ref{eq:normcphi}). We note that in the symmetric case $\bar{N}_a=\bar{N}_b$,
we recover the Gaussian dependence for $c(\phi)$ of Eq.(\ref{eq:gauss})
with $\Delta n \Delta \phi=1/2$.

We are now ready to calculate 
$\langle c_a^\dagger c_b\rangle^{\mbox{\scriptsize fix}}$
starting from Eq.(\ref{eq:respsit}).
The calculation closely follows the one in section \ref{sec:Observables1}.
In particular we use the key property Eq.(\ref{eq:shift}) to obtain:
\bea
\langle\psi(t)|c_a^{\dagger}c_b|\psi(t)\rangle &=&
{1\over\pi^2}|{\cal B}(t)|^2 |{\cal N}|^2
\int_{-\pi/2}^{\pi/2}d\phi\int_{-\pi/2}^{\pi/2}d\phi'
e^{-(\phi^2+{\phi^\prime}^2)\Delta n^2}
e^{i(\kappa-\alpha)(\phi-\phi')} \nonumber \\
               &{\tilde{N}\over 2}&
e^{-i[\phi+\phi'+2(D+v(\tilde{N})t)]}
{}_{\tilde{N}-1}\langle\phi'-\chi t|\phi\rangle_{\tilde{N}-1}.
\label{eq:jolie}
\eea
The phase factor $e^{i\kappa(\phi-\phi')}$ in the integrand varies
rapidly with $\phi-\phi'$ at the scale $1/\sqrt{N}$ when
$\bar{N}_b-\bar{N}_a$ is larger than $\sqrt{N}$.
For this reason 
we approximate the scalar product between the phase states 
$|\phi\rangle_{\tilde{N}}$ and $|\phi'\rangle_{\tilde{N}}$ by a
Gaussian $\exp(-\tilde{N}(\phi-\phi')^2/2)$ rather than by the
$\delta$ distribution of section \ref{sec:Observables1}.
This leads to the approximation
\begin{equation}
 {}_{\tilde{N}-1}\langle\phi^\prime-\chi t|\phi\rangle_{\tilde{N}-1}
 \simeq  (-1)^{q_0(\tilde{N}-1)}
e^{-(\tilde{N}-1)(\phi^\prime-\phi-\chi t+q_0\pi)^2/2}
\end{equation}
where the integer $q_0$ is chosen 
such that $-\pi/2<(\chi t-q_0\pi)\leq \pi/2$.
By extending the limits of integration over $\phi,\phi'$
to $\pm\infty$
in Eq.(\ref{eq:jolie}) we are then left with a double Gaussian integral
that can be calculated exactly. The result is quite involved but  
it can be simplified by using the condition (\ref{eq:condsigma}) and
Eq.(\ref{eq:condn}). 
We take the average over the stochastic realizations and we
use again Eq.(\ref{eq:condn}) to simplify the result.
We calculate
the normalization factor ${\cal B}(t)$:
\be
1 \simeq {1 \over \pi^2} |{\cal N}|^2 |{\cal B}|^2(t)
\left(2\pi \over 4 \Delta n^2\right)^{1/2}
   \left( 2 \pi \over \tilde{N}+\Delta n^2 \right)^{1/2} 
e^{-{1\over 2}(\kappa-\alpha)^2/(\tilde{N}+\Delta n^2)}
\label{eq:a_asym}        \;.
\ee

We finally obtain for the mean contrast of the interference fringes between
A and B as:
\bea
\langle c_a^{\dagger}c_b\rangle^{\mbox{\scriptsize fix}} &\simeq&
  e^{-\lambda t} e^{-2i v(N)t}  
 \sum_{q=0}^{+\infty} e^{-{1 \over 2} \Delta n^2 [(\chi t-q \pi)]^2}
            (-1)^{q(N-1)} \nonumber \\
 & \sum_{k=0}^{N/m-1} &
 {\tilde{N}\over 2} 
 e^{-i\kappa (\chi t-q \pi) {\tilde{N}-1 \over \Delta n^2+\tilde{N}-1}}
 \frac{1}{k!} [\lambda t U(t)]^k 
\label{eq:Yvan_asym}
\eea
where the function $U(t)$ is given by:
\be
U(t) = \frac{1}{\lambda}
\left( \lambda_b 
 \frac{e^{im \mu_b^\prime t/\hbar}-1}{im \mu_b^\prime t/\hbar} +
       \lambda_a
 \frac{e^{-im \mu_a^\prime t/\hbar}-1}{-im \mu_a^\prime t/\hbar} 
\right)
\label{eq:newU} \;.
\ee


\newpage
\section*{Figures}

\begin{figure}[htb]
\caption{\it Two BECs A and B in two non overlapping trapping
potentials. Some atoms can be let out of the condensates towards
a 50--50 atomic beam splitter.
The detection of the atoms in the output channels of the beam-splitter
realizes a measurement of the relative phase between the condensates. }
\label{fig:BECs}
\end{figure}

\begin{figure}[htb]
\caption{\it Collapses and revivals of $\langle c_a^\dagger c_b\rangle^{\mbox{\scriptsize fix}}$
for an initial phase state 
(a) without losses and
(b) in presence of 3-body
losses.
The calculation is performed
for ${}^{87}$Rb atoms in the $F=1,m_F=-1$ state
and for isotropic harmonic traps. The 3-body loss rate is
inferred from the experimental data of JILA.
The initial total number of atoms
is $N=301$, and the harmonic frequencies are
$\Omega_a/2\pi=\Omega_b/2\pi=500 Hz$.
Diamonds: numerical result with $2.5\times 10^4$ Monte Carlo wave functions.
Solid line: analytical result.}
\label{fig:acrb}
\end{figure}

\begin{figure}[htb]
\caption{\it Collapses and revivals of $\langle c_a^\dagger c_b\rangle^{\mbox{\scriptsize fix}}$
for an initial phase distribution broader than that of the phase state.
The initial total number of atoms
is $N=301$. The initial distribution for the difference in
the number of particles in the two condensates is Gaussian with
a standard deviation
$\Delta n=6$ and a vanishing mean (so that
$\bar{N}_a=\bar{N}_b$). The other parameters are as in fig.\ref{fig:acrb}b.
Diamonds: numerical result with $2.5\times 10^4$ Monte Carlo wave functions.
Solid line: analytical result.}
\label{fig:acrb2}
\end{figure}

\begin{figure}[htb]
\caption{\it Monte Carlo simulation of a multichannel detection experiment 
using the device in fig.\ref{fig:BECs} to
sample the relative phase distribution corresponding to the initial
state of Fig.\ref{fig:acrb2}. (a) Single realization of the multichannel detection: For each dephasing
$\beta_i=i\pi/10,i=0\ldots 9$
added to one of the input channels of the beam splitter, $p_+(\beta_i)$ 
(resp.\ $p_-(\beta_i)$)
particles are detected in the $+$ (resp.\ $-$)
output channel 
of the beam splitter with $p_+(\beta_i)+p_-(\beta_i)=p=20$.
The obtained integers $p_+(\beta_i)$ (diamonds) are fitted with the function
$k\cos^2(\phi_0-\beta)$ (solid line)
where $-\pi/2<\phi_0\leq\pi/2$ is the adjustable parameter, varying from one realization to the other.
(b) After 100 realizations of the multichannel detection (each starting with new
condensates): histogram for the obtained values of $\phi_0$.}
\label{fig:multich}
\end{figure}

\begin{figure}[htb]
\caption{\it Single realization relative phase distribution at $t=0$ and at the $2^{nd}$ revival
time $t=2\pi/\chi$ for three different Monte Carlo wave functions. 
The parameters are as in fig.\ref{fig:acrb2}. From upper left to lower right the wave functions
have experienced 0,3,1 and 0 quantum jumps respectively.} 
\label{fig:singreal}
\end{figure}

\begin{figure}[htb]
\caption{\it Relative phase probability distribution at $t=0$ and at the $2^{nd}$ revival 
time. The parameters are as in Fig.\ref{fig:acrb2}. Solid line: analytical prediction.
Diamonds: average of $2.5\times 10^4$ Monte Carlo wave functions.}
\label{fig:cphirev}
\end{figure}

\begin{figure}[htb]
\caption{\it Collapses and revivals of $\langle c_a^\dagger c_b\rangle^{\mbox{\scriptsize fix}}$ 
for a 10\% asymmetry
in the initial number of particles $\bar{N}_a$ and $\bar{N}_b$ in
the condensates $\bar{N}_a=135.5$ and $\bar{N}_b=165.5$,
leading to $\gamma_a \neq \gamma_b$, $\mu_a \neq \mu_b$.
The initial total number of atoms
is $N=301$. 
The initial distribution for the difference in the numbers of particles
$n$ in the condensates is Gaussian with a standard deviation 
$\Delta n=6$, and a non-vanishing mean value equal to 30. 
The other parameters are as in fig.\ref{fig:acrb}b.
Diamonds: numerical result with $2.5\times 10^4$ Monte Carlo wave functions.
Solid line: analytical result.}
\label{fig:asymm}
\end{figure}

\begin{figure}[htb]
\caption{\it Collapses and revivals of $\langle c_a^\dagger c_b\rangle^{\mbox{\scriptsize fix}} $ 
for an initial phase state with $N=301$ particles (solid line) and after an average
over $N$ with a Poisson distribution of parameter $\bar{N}=301$ (diamonds).
The effect of the average is mainly to suppress the odd revivals.
The parameters are as in fig.\ref{fig:acrb}b and the results are obtained from the analytical predictions.}
\label{fig:Poisson}
\end{figure}

\begin{figure}[htb]
\caption{\it Collision fluxes $\lambda^{(1)}$ (stars) and $\lambda^{(3)}$ 
(diamonds), due
to one-body and three-body collisions respectively, calculated as in
fig.\ref{fig:acrb}, and inverse of the first revival
time 
$1/t_{\mbox{\scriptsize rev}}=\chi/\pi$ (solid line) 
as a function of the total number of atoms. 
The trap frequency is $\Omega=2\pi\times
500 Hz$ in (a) and $\Omega= 2\pi\times 200 Hz$ in (b). 
The vertical dashed line for $\bar{N}=301$
in (a) represents the conditions of fig.\ref{fig:acrb}b.
$\lambda^{(1)}$ corresponds to a lifetime due to background gas collisions of 350 seconds.} 
\label{fig:feasib}
\end{figure}

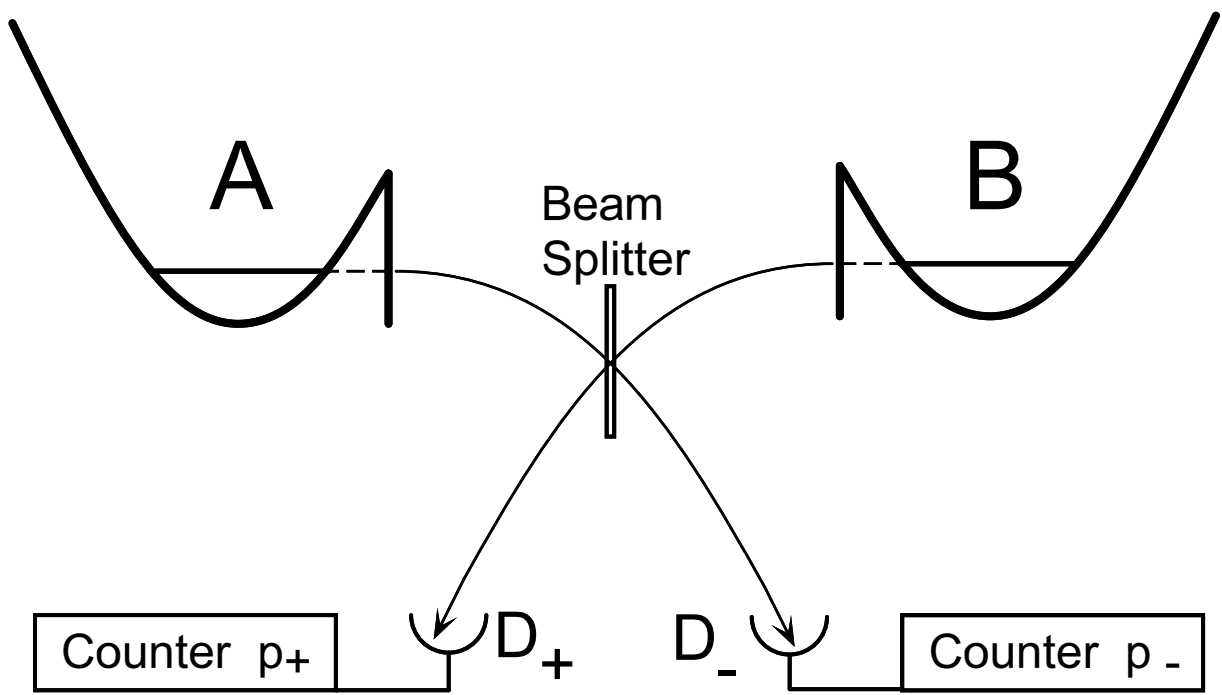
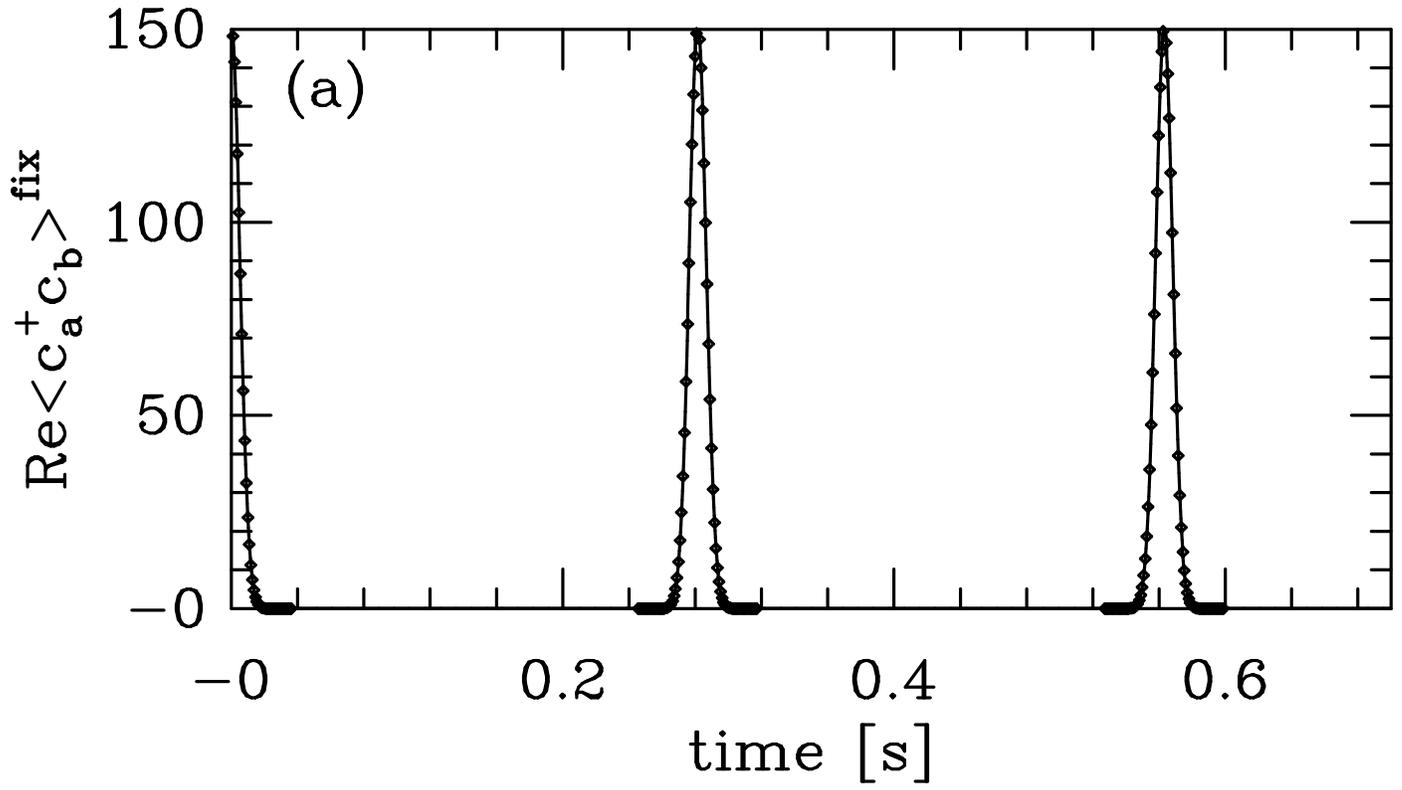
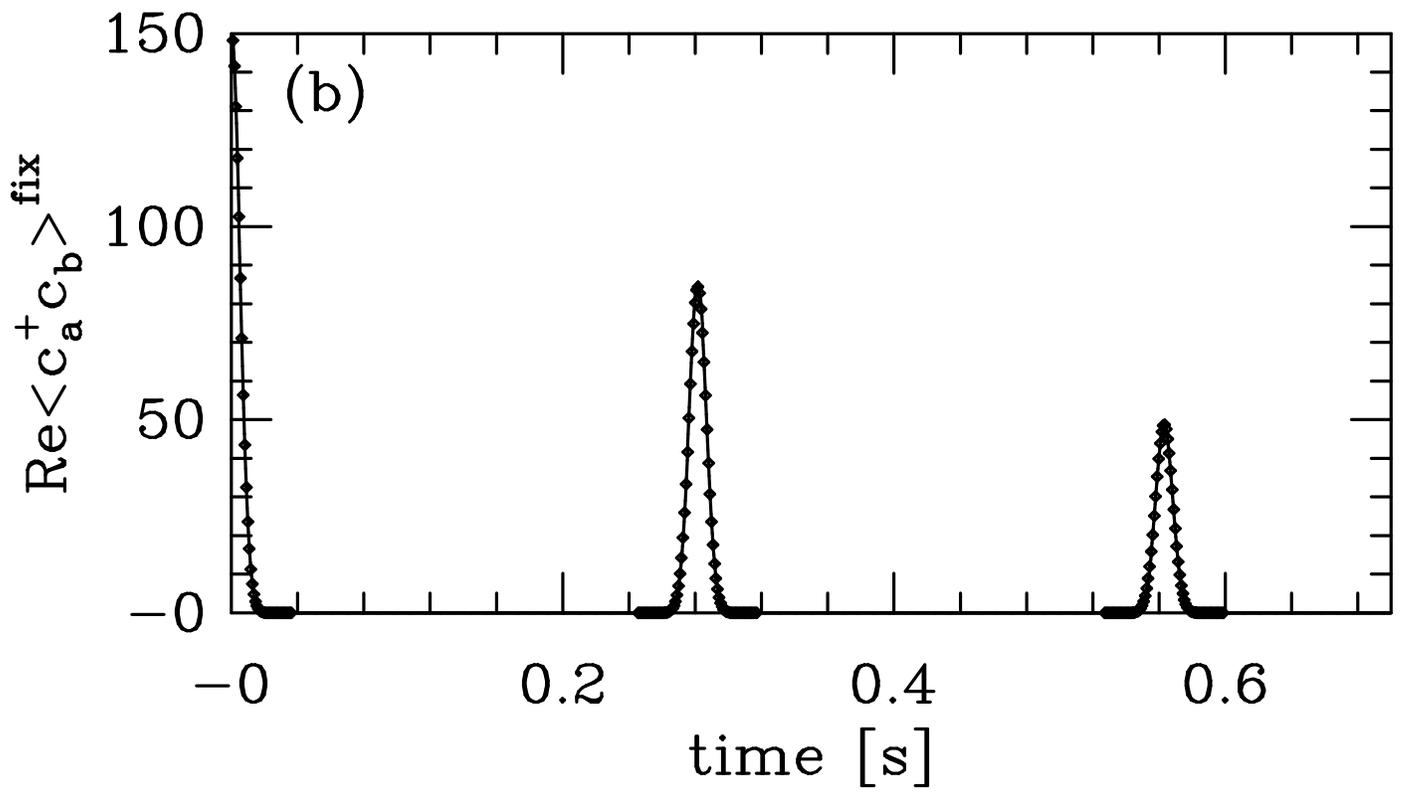
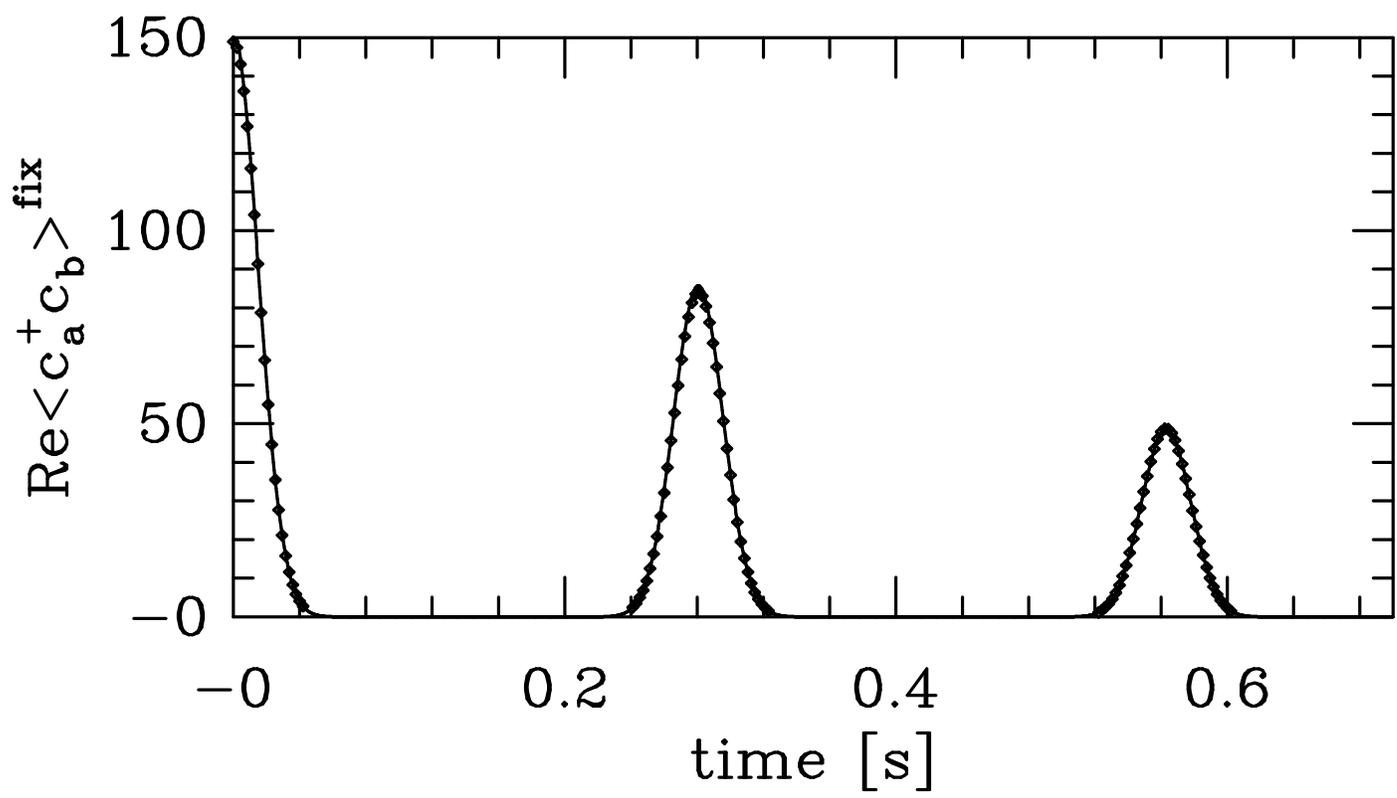
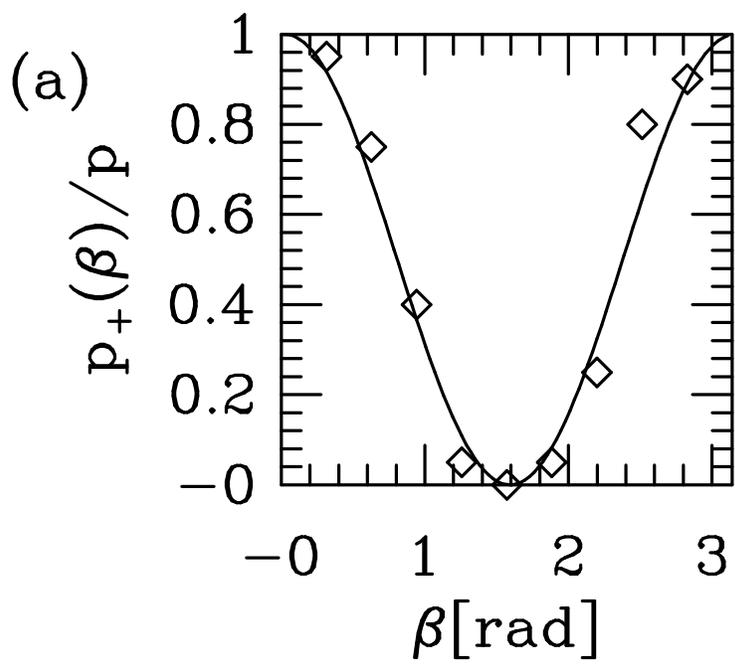
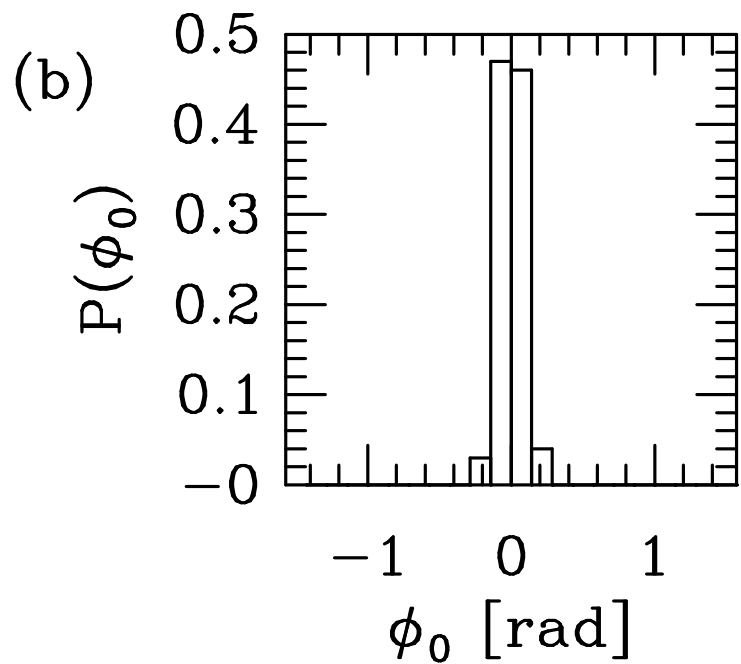
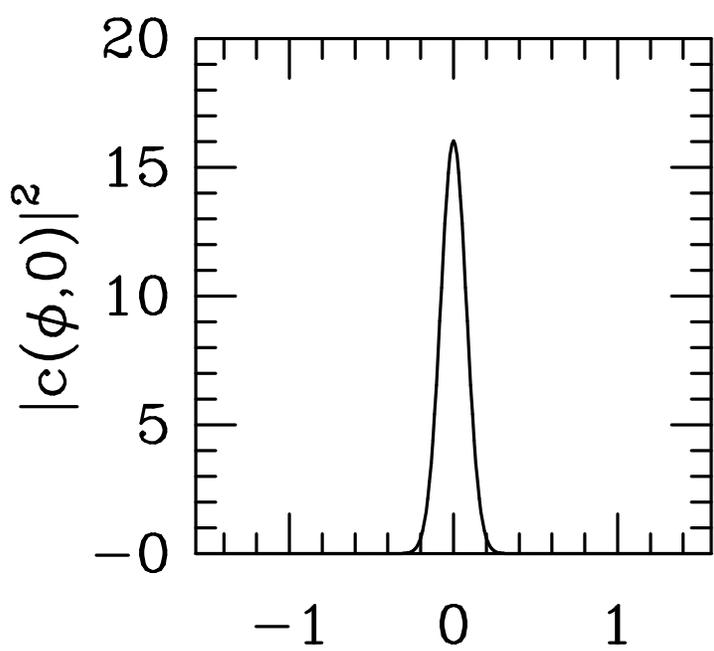
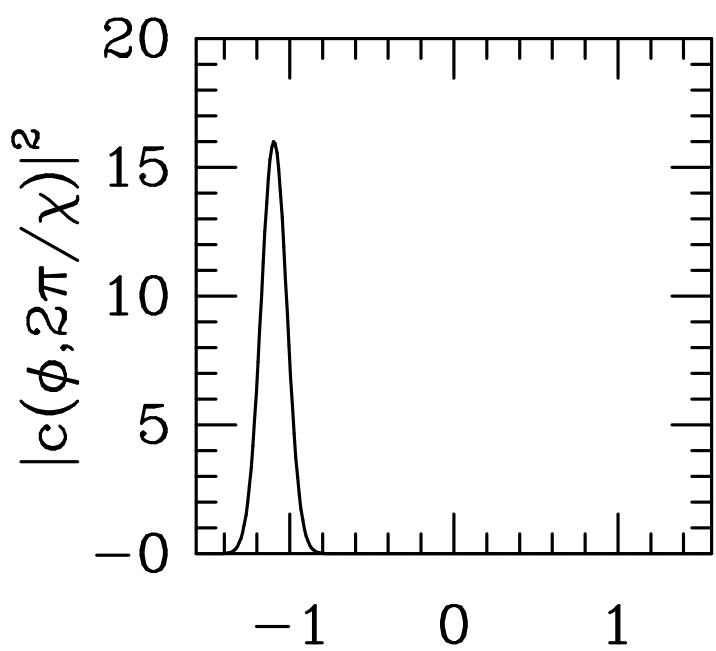
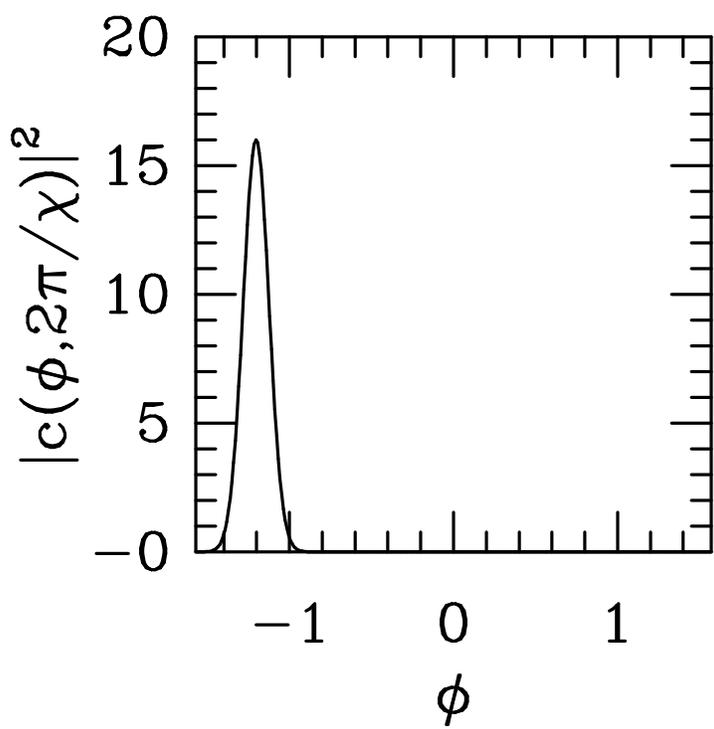
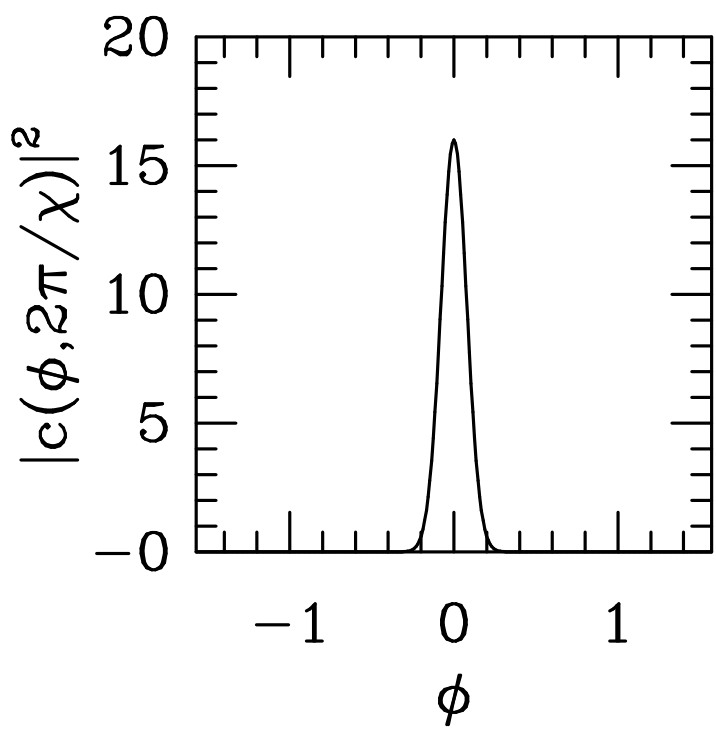
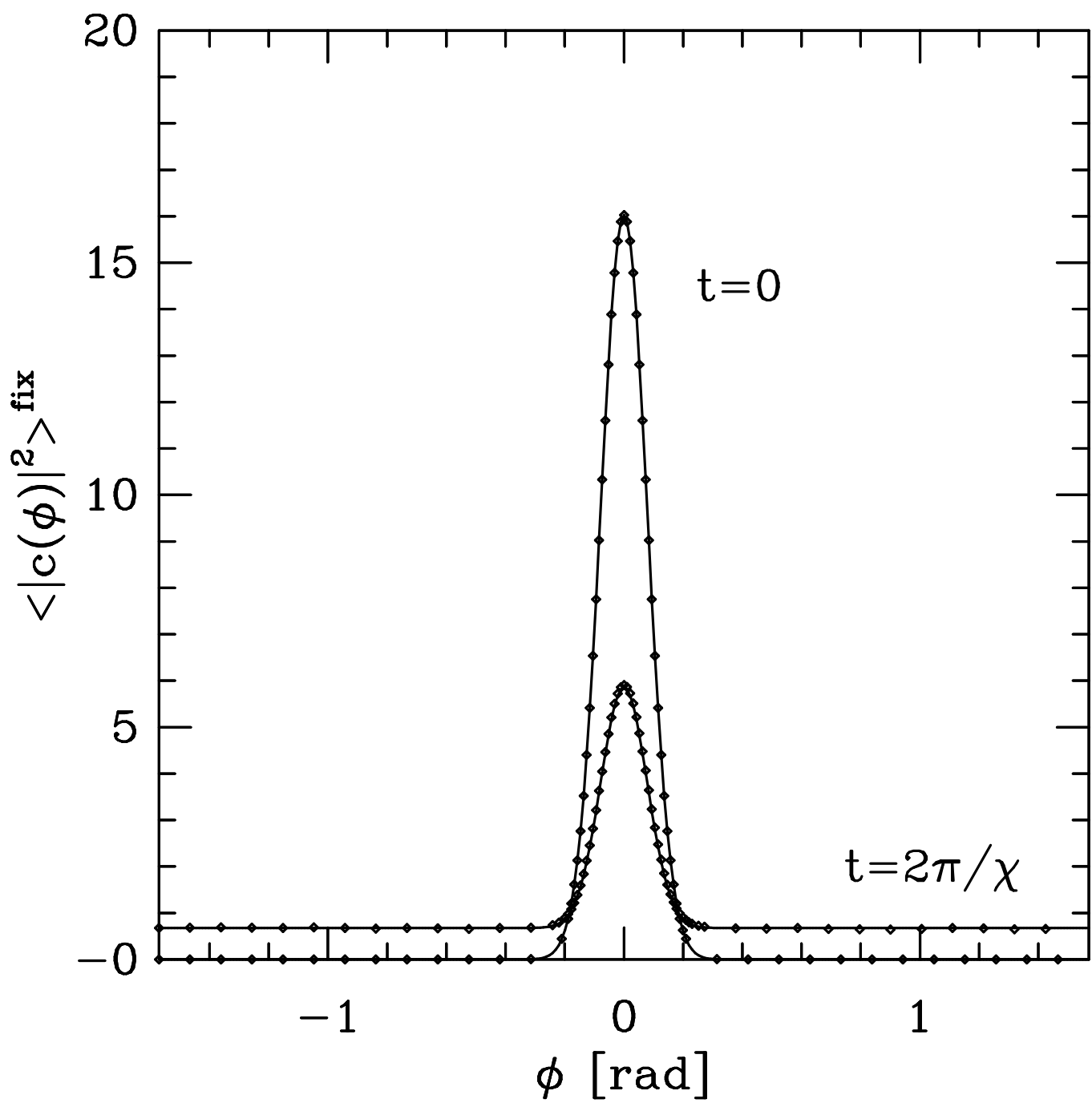
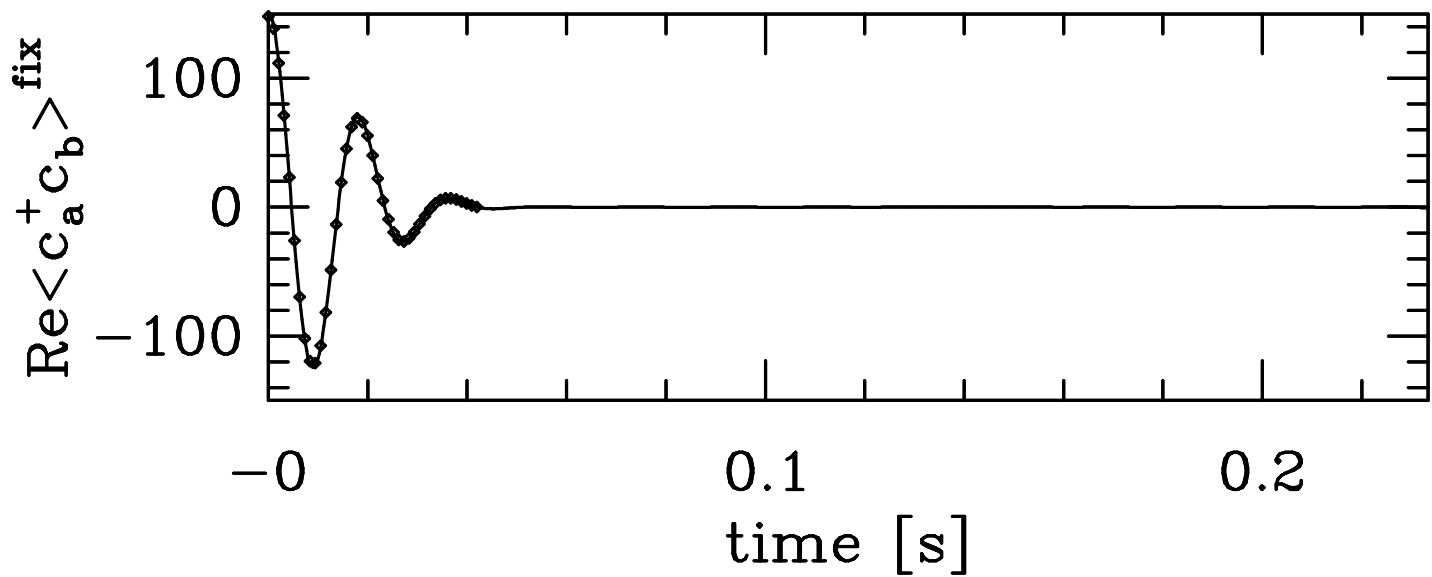
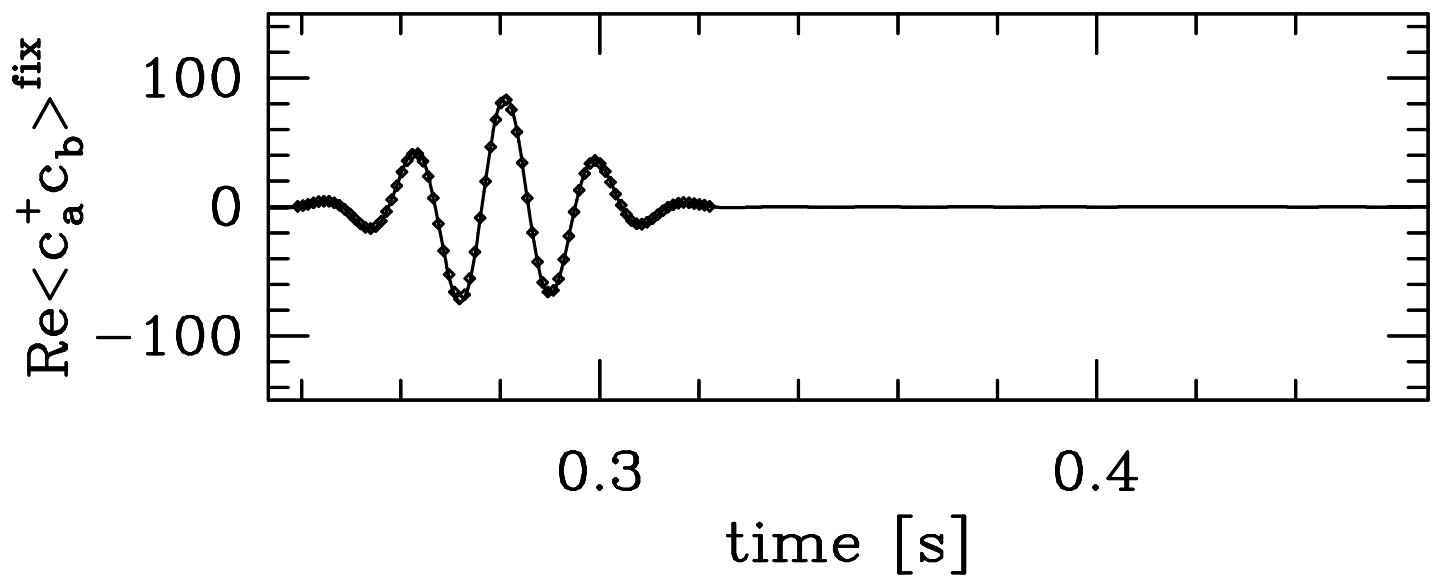
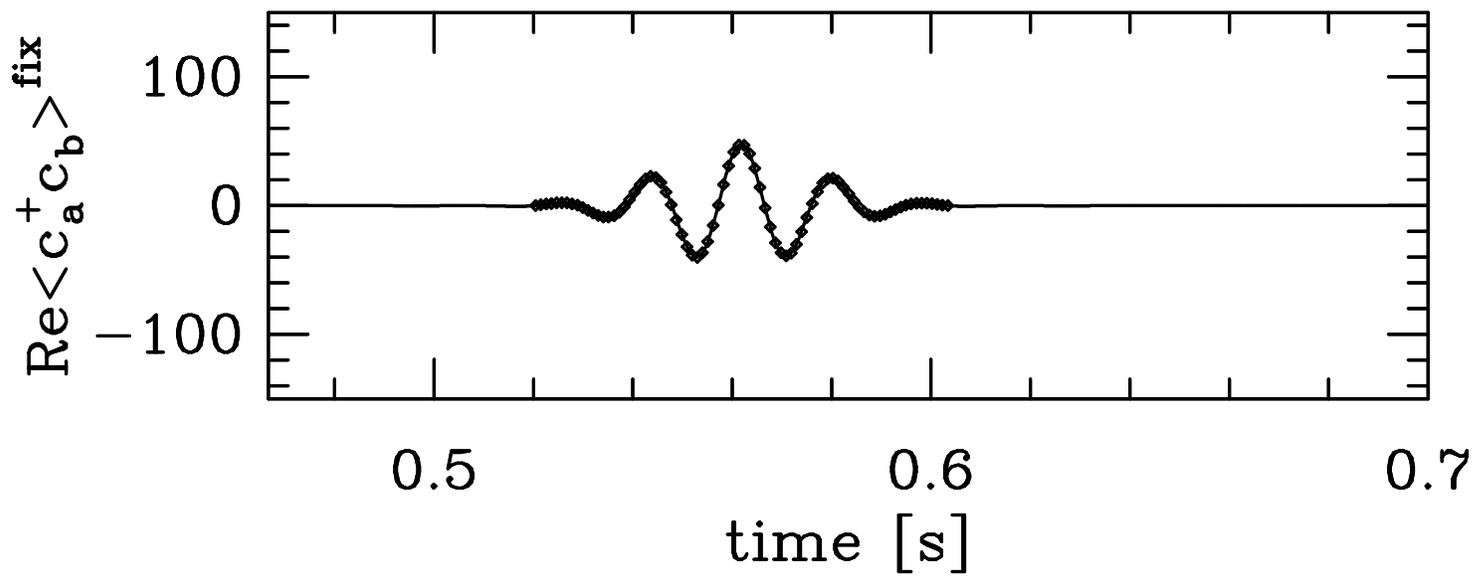
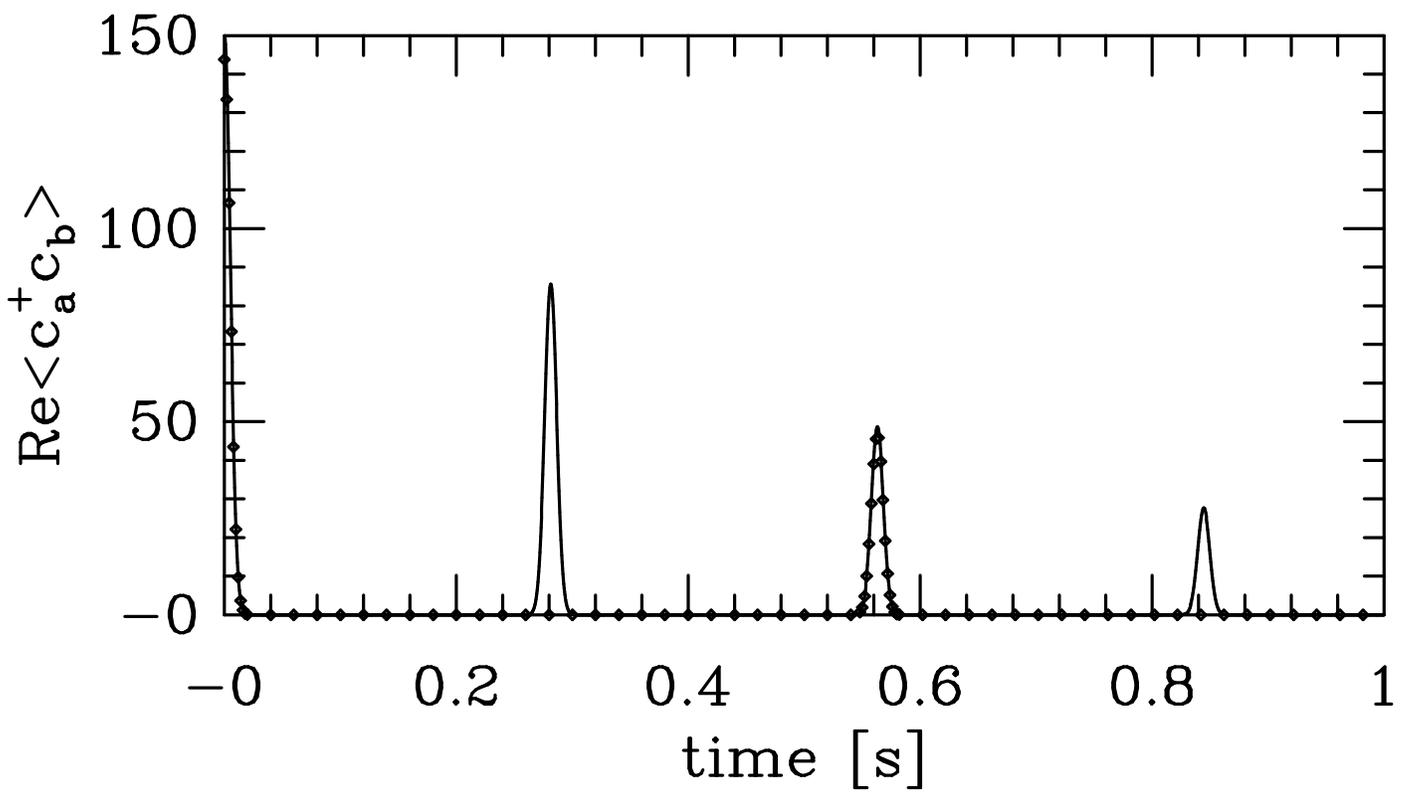
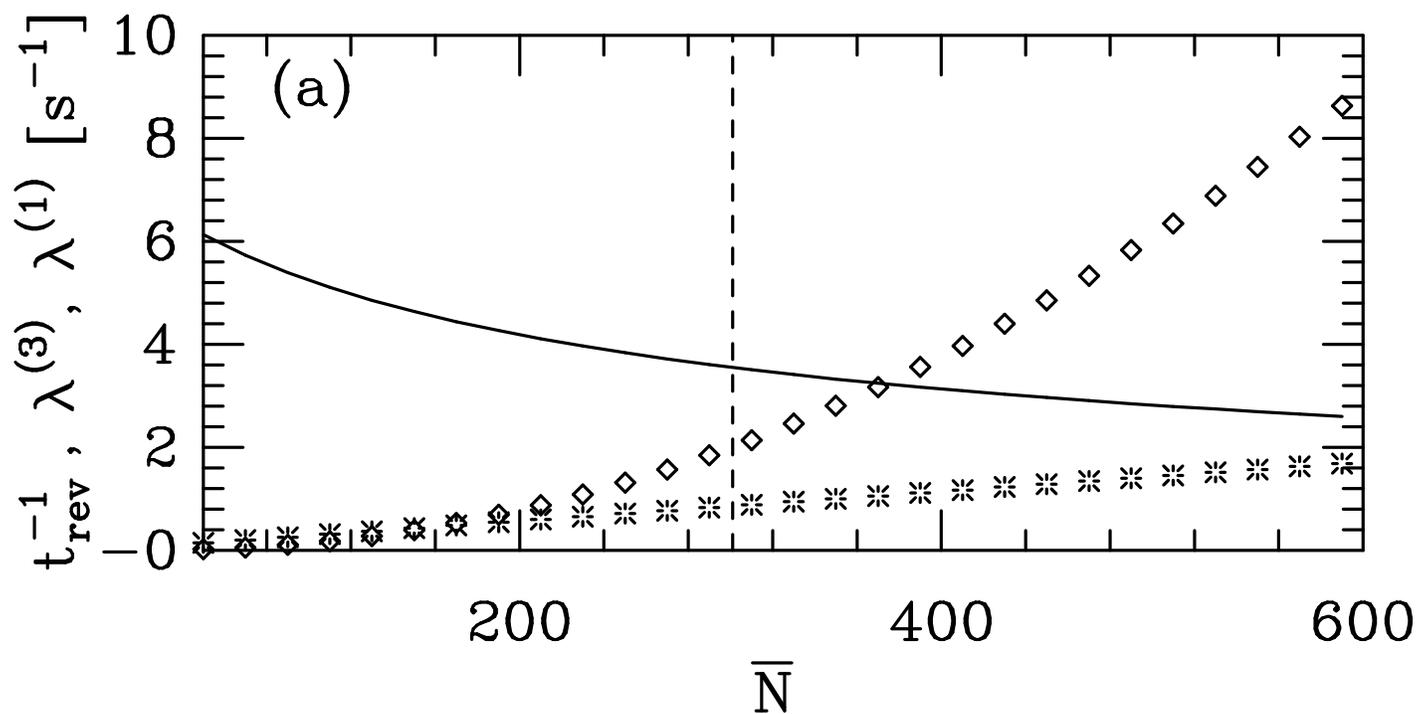
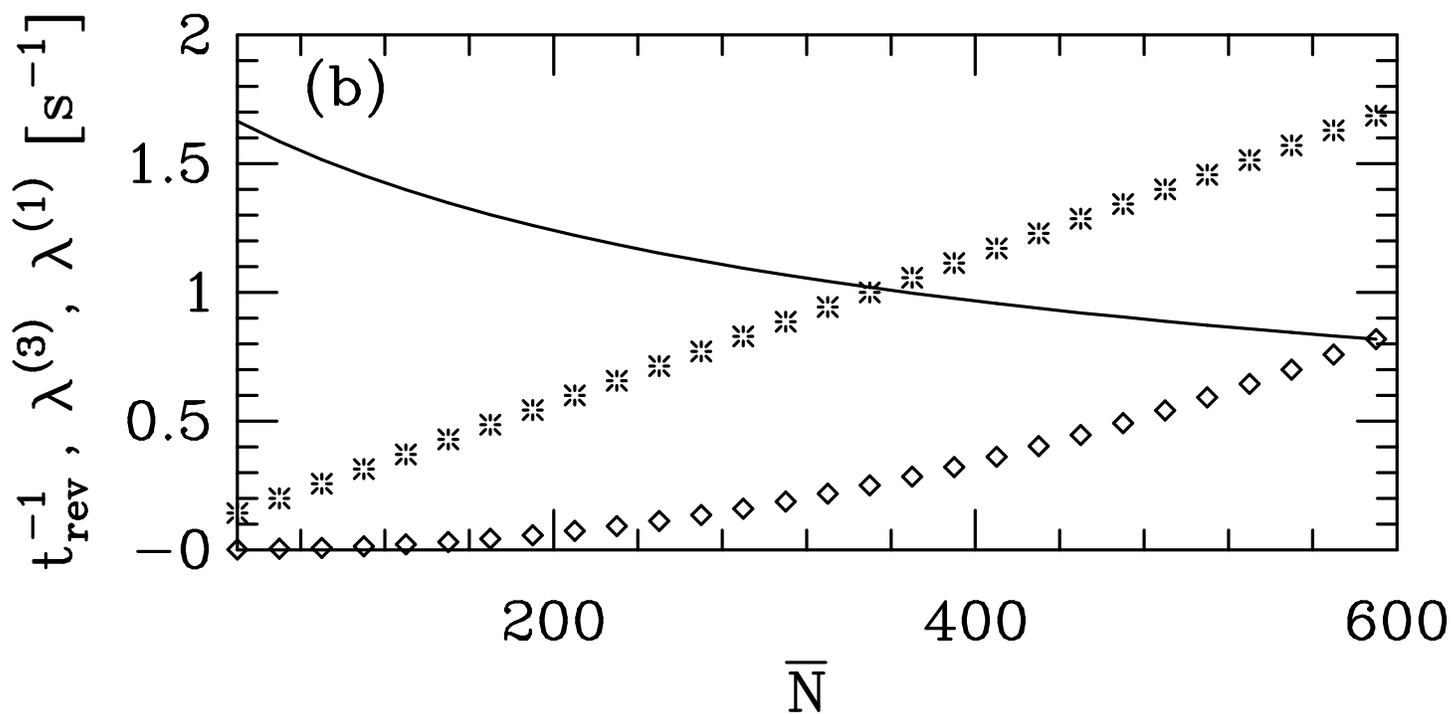
\end{document}